\documentclass[twocolumn,prb,aps,showpacs]{revtex4} 
\usepackage[dvips]{graphicx}
\usepackage{subfigure}

\usepackage{color}
\usepackage{amsmath}
\usepackage{dsfont}
\begin{document}
\title{Parquet decomposition calculations of the electronic self-energy}
\author{O.~Gunnarsson,$^1$ T.~Sch\"afer,$^2$ J.~ P.~ F.~LeBlanc,$^3$, J.~Merino,$^4$
G.~Sangiovanni,$^5$ G.~Rohringer,$^{2,6}$ and A.~Toschi$^2$ } 
\affiliation{
$^1$ Max-Planck-Institut f\"ur Festk\"orperforschung, Heisenbergstrasse 1, D-70569 Stuttgart, Germany \\ 
$^2$ Institute of Solid State Physics, Technische Universit\"at Wien, 1040 Vienna, Austria \\
$^3$ Department of Physics, University of Michigan, Ann Arbor, Michigan 48109, USA \\
$^4$ Departamento de F\'isica Te\'orica de la Materia Condensada, Condensed Matter Physics Center (IFIMAC) and
Instituto Nicol\'as Cabrera, Universidad Aut\'onoma de Madrid, Madrid 28049, Spain\\
$^5$ Institute of Physics and Astrophysics, University of W\"urzburg, W\"urzburg, Germany\\
$^6$Russian Quantum Center, Novaya street, 100, Skolkovo, Moscow region 143025, Russia} 

\begin{abstract}
The parquet decomposition of the self-energy into 
classes of diagrams, those associated with specific scattering processes, 
can be exploited for different scopes. In this work, the parquet decomposition
 is used to unravel the underlying physics of non-perturbative
numerical calculations. We show the specific example of
dynamical mean field theory (DMFT) and its cluster extensions (DCA)
applied to the Hubbard model at half-filling and with hole doping:
These techniques allow for a simultaneous determination of two-particle vertex functions and
self-energies, and hence, for an essentially ``exact'' parquet
decomposition at the single-site or at the cluster level.
Our calculations show that the 
self-energies in the underdoped regime 
are dominated by spin scattering processes, consistent with the
conclusions obtained by means of the fluctuation diagnostics approach
[Phys.\ Rev.\ Lett.\ {\bf 114}, 236402 (2015)]. 
However, differently from the latter approach, the parquet procedure displays
important changes with increasing interaction: Even for relatively
moderate couplings, well before the Mott transition, singularities appear in
different terms, with the notable exception of the predominant
spin-channel. We explain precisely how these singularities, which
partly limit the utility of the parquet decomposition, and - more generally - of parquet-based algorithms, are never found in the
fluctuation diagnostics procedure. Finally, by a more refined
analysis, we link the occurrence of the parquet singularities in our calculations to a progressive
suppression of charge fluctuations and the formation of an
RVB state, which are typical hallmarks of a pseudogap state in DCA. 
\end{abstract}
\date{\today} 


\maketitle 

\section{Introduction}\label{sec:1}

Traditionally, the electron self-energy is often determined via diagram 
expansion methods.\cite{Hedin,AGD} Diagrams to low order in the interaction strength 
can be calculated in perturbation theory. It may also be possible to sum 
certain classes of diagrams to infinite order. For instance, the lowest 
order diagram in the screened Coulomb interaction, the GW method,\cite{GW} 
gives reasonable results for moderately correlated systems, such 
as free-electron-like metals and semiconductors.\cite{Ferdi,Wilkins} 
Even in the case of a strongly correlated system like NiO certain aspects 
are described reasonably well, but, still, important parts of the physics are 
believed to be missing.\cite{NiO} Including the next order terms in such an
expansion can even lead to wrong analytical behavior.\cite{Minnhagen} Improving 
further in this respect, would require the consideration
of the contributions to the electron self-energy of different channels simultaneously, 
as it is done in FLEX,\cite{FLEX} functional renormalization group,\cite{fRGrev} 
or the parquet approximation.\cite{Bickers,PA_Janis,PA_Jarrell} Despite 
the ever increasing numerical workload of these schemes, they 
often do not improve upon the GW for the description of crucial aspects of correlated
systems. For instance, they also fail to capture the physics of the
Mott-Hubbard metal insulator transition, whose nature is intrinsically
non-perturbative.

To overcome these difficulties, completely different {\sl and}
non-perturbative methods, 
such as self-consistently 
embedded impurity/cluster algorithms like the dynamical mean field theory 
(DMFT)\cite{DMFTrev}, dynamical cluster approximation (DCA)\cite{Maier,Hettler} and
cellular DMFT (CDMFT)\cite{Kotliar} have been introduced, and
are now widely used. In such methods,  a cluster with a finite number
($N_c$) of atoms 
is embedded in a self-consistent host of noninteracting electrons. The cluster 
problem can be solved by diagonalization algorithms but for most cases Quantum 
Monte Carlo (QMC) methods are more efficient, e.g., in its Hirsch-Fye\cite{HirschFye} 
or continuous time (CT)\cite{CT} version. In this approach the only essential
approximation is the limitation to a finite cluster and the convergence of the 
results with $N_c$ can be checked systematically\cite{LeBlanc2015}. In cases where the Monte-Carlo 
sign problem is not serious, these methods can provide very reliable results 
for the electron self-energy. In the last years, also calculations of two-particle vertex
functions\cite{vertex_DCA,fourier,Kunes2011,Rohringer2012,Park2011,Hafermann2014,GangLi2016} 
became possible. This technical progress has a very high 
impact, because two-particle vertex functions are a crucial ingredient for
calculating\cite{DMFTrev,Maier} momentum- and frequency-dependent response
functions in DMFT and DCA, and also represent the building blocks for all 
multiscale extensions of DMFT\cite{DGA,DF,1PI,DMF2RG,TRILEX,FLEXDMFT} and DCA,\cite{Multiscale,DFparquet} 
aiming at including spatial correlations on all length
scales.\cite{Rohringer2011,Antipov2014,Otsuki2014,Schaefer2015A, Schaefer2015B,Pudleiner,Hirschmeier2015}

The purpose of this paper is, however, not to obtain new result for the self-energy 
with these novel schemes. In fact, at least within DMFT and DCA, the self-energy 
can be directly computed without the time consuming calculation of the two-particle
vertices. Our aims, here, are different: (i) to develop
methods that improve our physical interpretation of the
self-energy results in strongly correlated 
systems, and (ii) to understand how the correlated physics is actually
captured by diagrammatic approaches beyond the perturbative regime.

We do this by applying a parquet-based diagrammatic 
decomposition to the self-energy. Specifically, we use the DMFT and DCA results for this
parquet decomposition, thus avoiding
{\sl any} perturbative approximation for the vertex. 
We apply the method to the Hubbard model on cubic (three
dimensional, $3d$) and square (two-dimensional, $2d$) lattices. In
these cases, quite a bit is already known about the physics, 
which, to some extent, allows for a check of our methodology.

We recall briefly here, that in the parquet schemes two-particle diagrams are 
classified according to whether they are two-particle reducible (2PR) in a certain channel, i.e., 
whether a diagram can be split in two parts by only cutting two Green's functions, 
or are fully irreducible at the two-particle level (2PI). Diagrams reducible in a 
particular channel can then be related to specific physical processes. 
Specifically, we obtain three classes of reducible diagrams, longitudinal ($ph$)
and transverse ($\bar{ph}$) particle-hole diagrams and particle-particle ($pp$) 
diagrams. Because of the electronic spin, the particle-hole diagrams
can be rearranged, more physically, in terms
of spin (magnetic) and charge (density) contributions, while for $pp$
 the $\uparrow \downarrow$ term (essential for the singlet pairing) will be explicitly kept.

In this work, we compute explicitly the parquet equations, Bethe-Salpeter equations and the equation of motion (EOM) which 
relate the vertices in the different channels to each other and to the
self-energy, by using the 2PR and 2PI vertices 
of the DMFT and  DCA calculations. Hence, apart from statistical errors, we  get an ``exact'' diagrammatic 
expansion of the self-energy of our DMFT ($N_c=1$) or DCA ($N_c > 1$)
clusters.  Since, within the parquet formalism,  the physical
processes are automatically associated to  the different scattering
channels,  our calculations can be exploited to extract an unbiased physical interpretation 
of our DMFT and DCA self-energies and to investigate the structure of the
Feynman diagrammatics beyond the perturbative regime. 
We note here that, from the merely conceptual point of
view, the parquet decomposition is the most ``natural'' route to
disentangle the physical information encoded in self-energies and
correlated spectral functions. The parquet procedure can be compared, e.g., to the
recently introduced fluctuation diagnostics\cite{FluctDiag} approach,
which also aims at extracting the underlying physics of a given self-energy:
In the fluctuation diagnostics the quantitative information
about the role played by the different physical processes is extracted by studying
the different representations (e.g., charge, spin, or
particle-particle), in which the EOM for the self-energy, and
specifically the full two-particle scattering amplitude, can be written. 
Hence, in this respect, the parquet decomposition provides a more
direct procedure, because it does not require
any further change of representation for the momentum, frequency, spin
variables, and can be readily analyzed at once, provided that the
vertex functions have been calculated in an channel-unbiased way. 
However, as we will discuss in this work, the parquet decomposition
presents also disadvantages w.r.t. the fluctuation
diagnostics, because (i)  it requires working with 2PI vertices, which
makes the procedure somewhat harder
from a numerical point of view, and  (ii) it faces intrinsic instabilities for increasing interaction values.

By applying this procedure to the 2d Hubbard model at intermediate values of 
$U$ (of the order of half the bandwidth), we find large contributions
from spin-fluctuations. This is consistent with
a common belief that ${\bf Q}=(\pi,\pi)$ spin fluctuations are very important
for the physics, as well as with the fluctuation diagnostics results.\cite{FluctDiag} For the 3d Hubbard model similar physics was first proposed 
by Berk-Schrieffer.\cite{Berk} Later spin fluctuations have been proposed 
to be important for the 2d Hubbard model and similar models by many 
groups.\cite{Scalapinospin,spinfluctuations,Bulut,Haule} 
We note, however, that the contributions of the other channels to the
parquet decomposition are not small by themselves. Rather, the other
(non-spin) channel contributions to $\Sigma({\bf k}, i\nu)$ appear to
play the role of ``screening'' the electronic scattering originated by
the purely spin-processes. The latter would lead, otherwise, to a
significant overestimation of the electronic scattering rate. At
larger values of $U$ the parquet decomposition starts displaying
strong oscillation at low-frequencies in all its term, {\sl but} the
spin contribution. Physically, this might be an indication that the spin
fluctuations also predominate in the non-perturbative regime, where,
however, the parquet distinction among the remaining (secondary) channels 
loses its physical meaning. The reason for this can be traced back to
the occurrence of singularities in the generalized susceptibilities of
these (secondary) channels. Such singularities are reflected
in the corresponding divergencies of the two-particle
{\sl irreducible} vertex functions, recently discovered in the DMFT
solution of the Hubbard and Falicov-Kimball 
models\cite{divergence,Janis2014,Yang,Kozik2015,Stan2015,Rossi2015,Ribic2016}. Here we extend
the study of their origin and generalize earlier results \cite{divergence}
to DCA. We discuss the relation of these singularities to the resonance valence
bond RVB\cite{Liang88} character of the ground-state, the pseudogap and the
suppression of charge fluctuations for large values of $U$.

Our results are relevant also beyond the specific problem of the
physical interpretation of the self-energy. In fact, the parquet
decomposition can be also used to develop new quantum many-body schemes.
Wherein some simple approximation might be introduced for the irreducible diagrams
that are considered to be particularly fundamental. The parquet
equations are then used to calculate the reducible diagrams. In our results,
however, for strongly correlated systems the contribution to the self-energy from
the irreducible diagrams diverges for certain values of $U$ both in
DMFT and DCA. This makes the derivation of good
approximations for these diagrams for strongly correlated systems rather challenging. It remains, however, an interesting question
if the parquet decomposition can be modified in such a way that these
problems are avoided.

\begin{figure*}[t!]
{\rotatebox{0}{\resizebox{18cm}{!}{\includegraphics {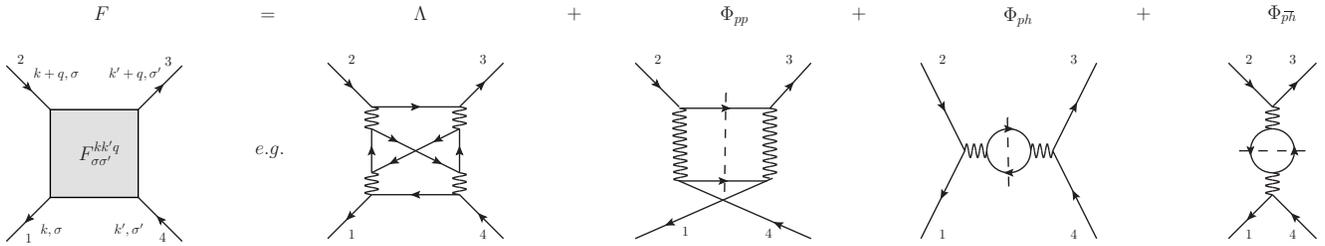}}}}
\vspace{2mm}
\caption{\label{fig:2.1} Two-particle vertex function $F$ (left) and
  its diagrammatic parquet decomposition (right), exemplified by the
  corresponding lowest order diagrams beyond the bare $U$. The
  (two-particle) cutting procedure indicating the two-particle reducibility of the last three terms is shown
by the dashed lines.}
\end{figure*}

The scheme of the paper is the following. In Sec.~\ref{sec:2} we present 
the formalism relating the vertex function to generalized two-particle 
response functions as well as the parquet decomposition of the vertex 
function. We also briefly describe the model and the calculation method. 
In Sec.~\ref{sec:3} we show results from the parquet decomposition and its 
behavior for intermediate and large $U$. In Sec.~\ref{sec:4} the behavior of
the generalized susceptibility is discussed, and the origin of singularities 
in the generalized charge response function is shown. In Sec.~\ref{sec:5} 
we discuss the relation of these singularities to the RVB character of the 
system, the pseudogap and the suppression of charge fluctuations. Sec.~\ref{sec:6}
is devoted to our conclusions.

\begin{figure}
{\rotatebox{0}{\resizebox{5.5cm}{!}{\includegraphics {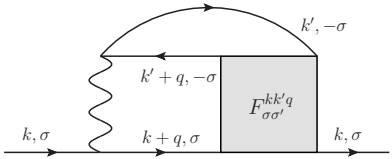}}}}
\vspace{2mm}
\caption{\label{fig:2.2} Diagrammatic representation of the self-energy $\Sigma$ in terms of the two-particle vertex function (Schwinger-Dyson equation of motion).}
\end{figure}

\section{Formalism, model and  method}\label{sec:2}

We first discuss the vertex function, following the notations of Rohringer {\it et al.}\cite{Rohringer2012} and Gunnarsson {\it et al.}\cite{FluctDiag}
We introduce the generalized susceptibility for finite temperature $T=1/\beta$, using the Matsubara formalism
\begin{eqnarray}\label{eq:2.1}
&&\chi_{\sigma \sigma'}(k;k';q)          
=\int_0^{\beta}d\tau_1 \int_0^{\beta}d\tau_2 \int_0^{\beta}d\tau_3  \nonumber  \\
&& \times e^{-i[\nu\tau_1-(\omega+\nu)\tau_2+(\omega+\nu')\tau_3]}    \\
&& \times \langle T_{\tau}[c^{\dagger}_{{\bf k}\sigma}(\tau_1) c_{{\bf
    k}+{\bf q}\sigma}(\tau_2) c^{\dagger}_{{\bf k}'+{\bf
    q}\sigma'}(\tau_3) c_{{\bf k}'\sigma'}]\rangle  \nonumber \\ 
&& - \beta \, g_{\sigma}(k)g_{\sigma'}(k') \, \delta_{q=0}.\nonumber
\end{eqnarray}
Here we use the condensed notations $q=({\bf Q},\omega)$ and $k=({\bf
  K},\nu)$, where ${\bf Q}$ and ${\bf K}$ are (cluster) wave vectors and 
$\omega$ and $\nu$ are Matsubara boson and fermion frequencies, respectively. 
We have also introduced a Green's function $g_{\sigma}(k)\equiv g_{\sigma}({\bf K},\nu)$
\begin{equation}\label{eq:2.2}
g_{\sigma}(k)=-\int_0^{\beta} d\tau e^{i\nu \tau}\langle c_{{\bf K}\sigma}^{\phantom \dagger}(\tau)
c_{{\bf K}\sigma}^{\dagger} \rangle,
\end{equation}
where $c_{{\bf K}\sigma}^{\dagger}$ creates an electron with the wave vector 
${\bf K}$ and spin $\sigma$ and $\langle .. \rangle$ is the thermodynamical average.
From $\chi$, and specifically from its connected part, we obtain the full two-particle vertex $F$:
\begin{eqnarray}\label{eq:2.3}
&&\chi_{\sigma\sigma'}(k;k';q)=-\beta g_{\sigma}(k)g_\sigma(k+q)\delta_{kk'}\delta_{\sigma\sigma'} \\
&&-g_{\sigma}(k) g_{\sigma}(k+q)F_{\sigma\sigma'}(k;k';q)g_{\sigma'}(k') g_{\sigma'}(k'+q). \nonumber
\end{eqnarray}
The vertex function $F$ is shown diagrammatically in Fig.~\ref{fig:2.1}, and it can be interpreted, physically, as the scattering rate amplitude between two added/removed electrons. 
Within the parquet formalism all diagrams contributing to $F$ are divided in two classes: Either they can be split in two parts by cutting two internal Green's function lines (two particle reducibility: 2PR), or they cannot (two-particle irreducibility: 2PI) . 

\begin{figure*}[t]
\subfigure[]{\includegraphics[width=0.3\linewidth]{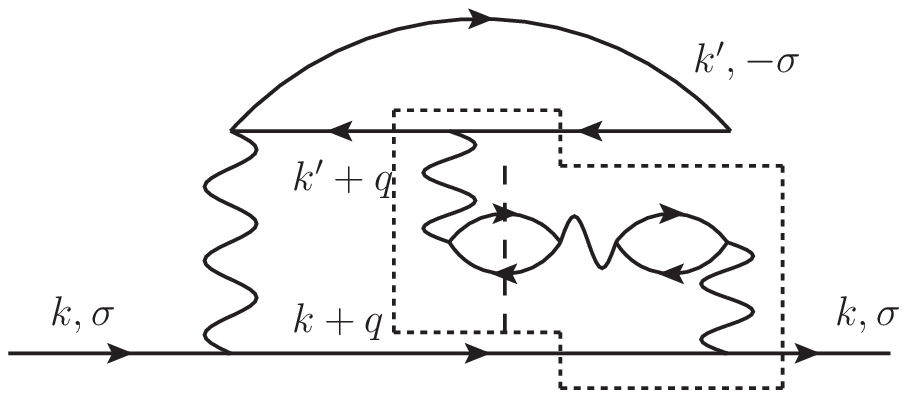}}
\subfigure[]{\includegraphics[width=0.3\linewidth]{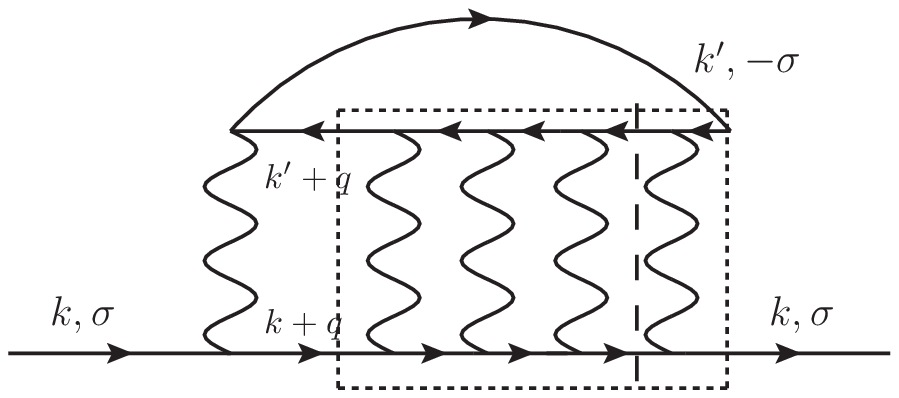}}
\subfigure[]{\includegraphics[width=0.3\linewidth]{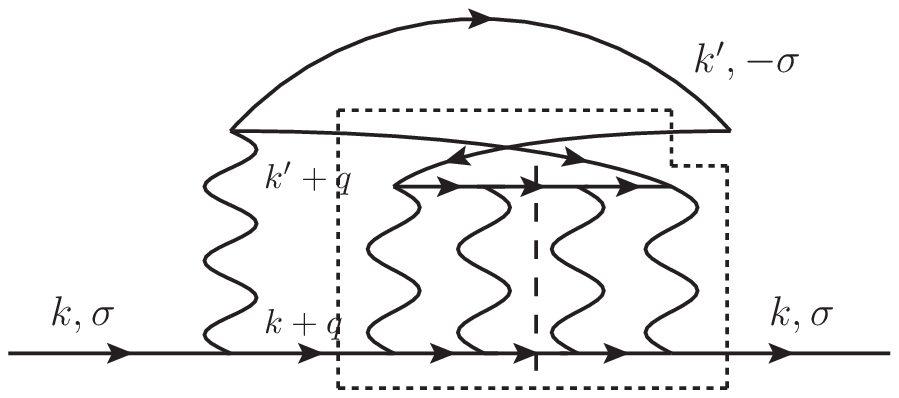}}
\caption{Examples of diagrams for the self-energy, with some explicit representations 
of the two-particle vertex function, enclosed by a dashed line. The 
dashed line shows how the vertex can be separated in two parts by cutting 
two Green's functions. According to the rules in Sec.~\ref{sec:2}, the 
diagrams are $ph$ (a), $\bar{ph}$ (b) and $pp$ (c) diagrams.}
\label{fig:3.0}
\end{figure*}

Moreover, because we are considering two-particle processes, whose
diagrams have (altogether) four external lines, a
finer classification can be performed for the 2PR diagrams. As exemplified by the diagrams
on the right-hand side of Fig.~\ref{fig:2.1}, we can further distinguish among the cases, where, in the cutting-procedure, (i) lines 1 and 3 are separated from 2 and 4, which corresponds to particle-particle 
($pp$) reducibility, (ii) lines 1 and 2 are separated from 3 and 4, i.e., longitudinal particle-hole 
($ph$) reducibility, and, eventually, (iii) lines 1 and 4 are separated from lines 2 and 3, i.e. transverse 
particle-hole ($\bar{ph}$) reducibility.  $F$ can then be written as a sum of these types of contribution
\begin{equation}\label{eq:2.4}
F=\Lambda+\Phi_{pp}+\Phi_{ph}+\Phi_{\bar{ph}},
\end{equation}
where $\Lambda$ contains the pure 2PI contributions and the functions 
$\Phi$ describe the 2PR contributions in all different 
channels, as diagrammatically represented in Fig.~\ref{fig:2.1}: This is
the {\sl parquet decomposition} of the scattering amplitude $F$. 

Finally, because of the electron spin,  it is convenient to treat the $ph$ channel 
by introducing generalized  charge ($ch$) and spin ($sp$) susceptibilities 
\begin{eqnarray}\label{eq:2.5}
&&\chi_{ch}(k;k';q)=\chi_{\uparrow\uparrow}(k;k';q)+ \chi_{\uparrow\downarrow}(k;k';q) \\
&&\chi_{sp}(k;k';q)=\chi_{\uparrow\uparrow}(k;k';q)- \chi_{\uparrow\downarrow}(k;k';q) \nonumber
\end{eqnarray}
We then define the quantities $\Gamma_d$ and $\Gamma_m$ which contain 
the diagrams of $F$ which are irreducible in the density and magnetic 
channels, respectively
\begin{equation}\label{eq:2.6}
\Gamma_{ch,sp}=\beta^2(\chi_{ch,sp}^{-1}-\chi_0^{-1}),
\end{equation}
where $\chi_0$ is the generalized bare susceptibility, 
being a product of two interacting Green's function.
The $\chi$'s are treated as matrices in $k$ and $k$' and $\Gamma$ 
can be calculated for one $q$ at a time. We also define the reducible 
quantities $\Phi_{ch,sp}$ via the Bethe-Salpeter equations
\begin{eqnarray}\label{eq:2.7}
&&\Phi_{ch,sp}=F_{ch,sp}-\Gamma_{ch,sp} \\
&&\Phi_{ph \uparrow \downarrow}=F_{\uparrow \downarrow}-{1\over 2}(\Gamma_{ch}-\Gamma_{sp}), \nonumber
\end{eqnarray}
and the parquet equations\cite{Rohringer2012}:
\begin{eqnarray}\label{eq:2.8}
&&\Lambda_{\uparrow\downarrow}(k,k',q)={1\over 2}[\Gamma_{ch}(k,k',q)-\Gamma_{sp}(k,k',q)] \nonumber \\
&&+\Phi_{sp}(k,k+q,k'-k) -\Phi_{pp}(k,k',k+k'+q) \\ \nonumber
\end{eqnarray}

By using the (Schwinger-Dyson) equation of motion, the electronic self-energy $\Sigma$ can 
be expressed in terms of two-particle vertex function:
\begin{eqnarray}\label{eq:2.9}
&&\Sigma(k)-{Un\over 2}            \\
&&=-{U\over \beta^2N_c}\sum_{k',q}F_{\uparrow \downarrow}(k,k',q)    
g(k')g(k'+q)g(k+q)  \nonumber
\end{eqnarray}
where $g=g_\uparrow=g_\downarrow$ (because of SU(2)-symmetry), $N_c$ is the number of ${\bf K}$-points. This is shown schematically 
in Fig.~\ref{fig:2.2}.

The equation of motion for $\Sigma$ is a well-known, general relation of many-body theory with a two-particle 
interaction. However, valuable information may be obtained by inserting  in Eq.~(\ref{eq:2.9}) the parquet decomposition of Eq.~(\ref{eq:2.4}) and, in particular, its specific expression for $F_{\uparrow \downarrow}(k,k',q)$:

\begin{eqnarray}\label{eq:2.4bis}
&& F_{\uparrow\downarrow}(k,k'\!,q)\!=\! \Lambda_{\uparrow\downarrow}(k,k'\!,q) \!+\! \Phi_{pp,\uparrow\downarrow}(k,k',k\!+\!k'\!+\!q) \\ 
&& +  \frac{1}{2}\Phi_{ch}(k,k',q)-\frac{1}{2}\Phi_{sp}(k,k',q) - \Phi_{sp}(k, k+q, k'-k) \nonumber
\end{eqnarray}

This way, {\sl after} all internal summations are performed, the expression for $\Sigma$ is naturally split in four terms:
\begin{equation}\label{eq:2.9bis}
\Sigma = \tilde{\Sigma}_{\Lambda}+\tilde{\Sigma}_{pp}+\tilde{\Sigma}_{ch}+\tilde{\Sigma}_{sp}
\end{equation}
evidently matching the corresponding 2PI and 2PR terms of Eq.~(\ref{eq:2.4bis}): 
This represents the {\sl parquet decomposition} of the self-energy. In fact, the four terms in Eq.~\ref{eq:2.9bis} describe the contribution of the different channels (pp, charge, spin), as well as of the 2PI scattering processes, to the self-energy. Since each scattering channel is associated with definite physical processes, Eq.~(\ref{eq:2.9bis}) can be exploited, in principle, for gaining a better understanding of the physics underlying a given self-energy calculation.


\begin{figure*}[t!]
{\rotatebox{-90}{\resizebox{6.0cm}{!}{\includegraphics {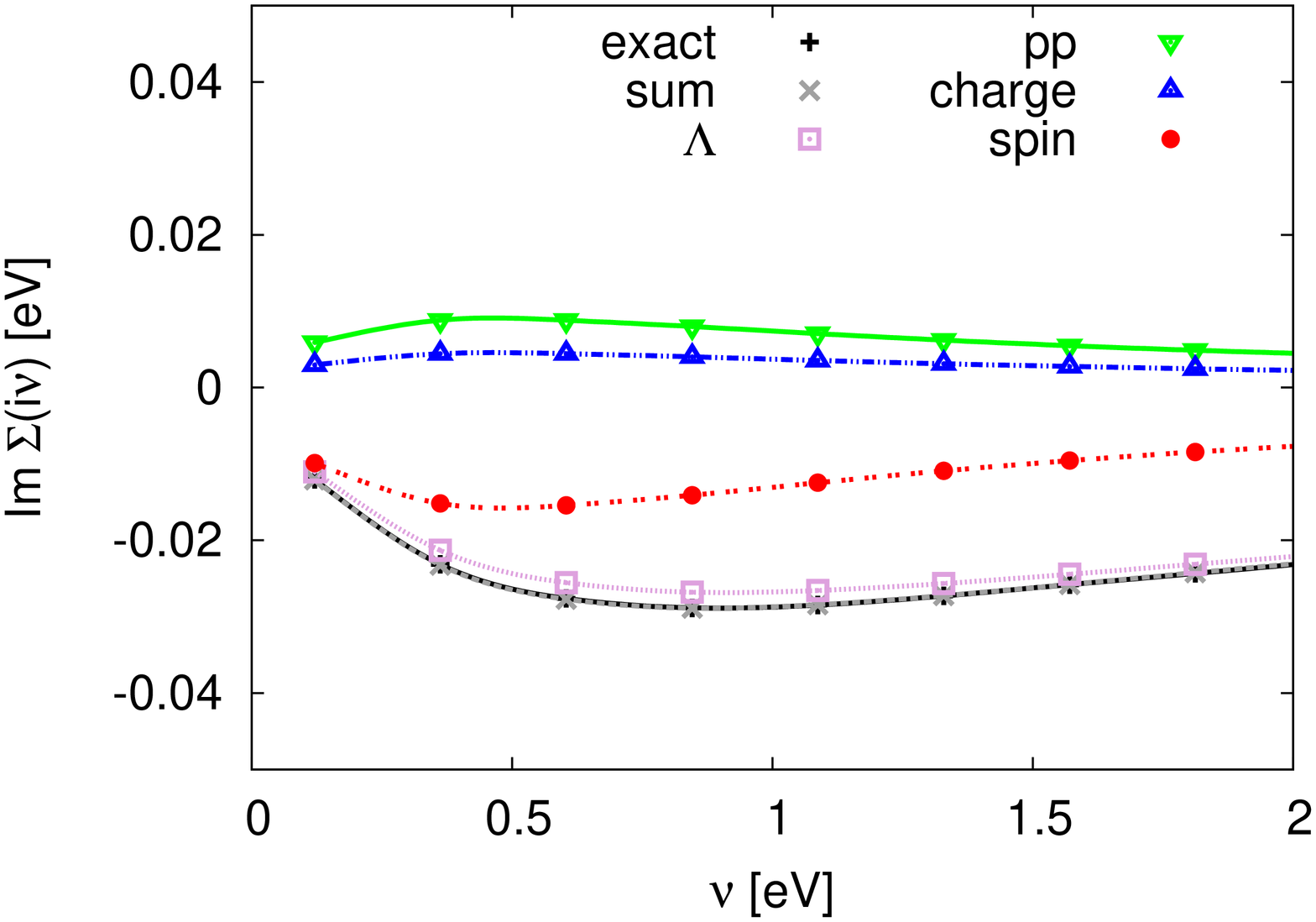}}}}
{\rotatebox{-90}{\resizebox{6.0cm}{!}{\includegraphics {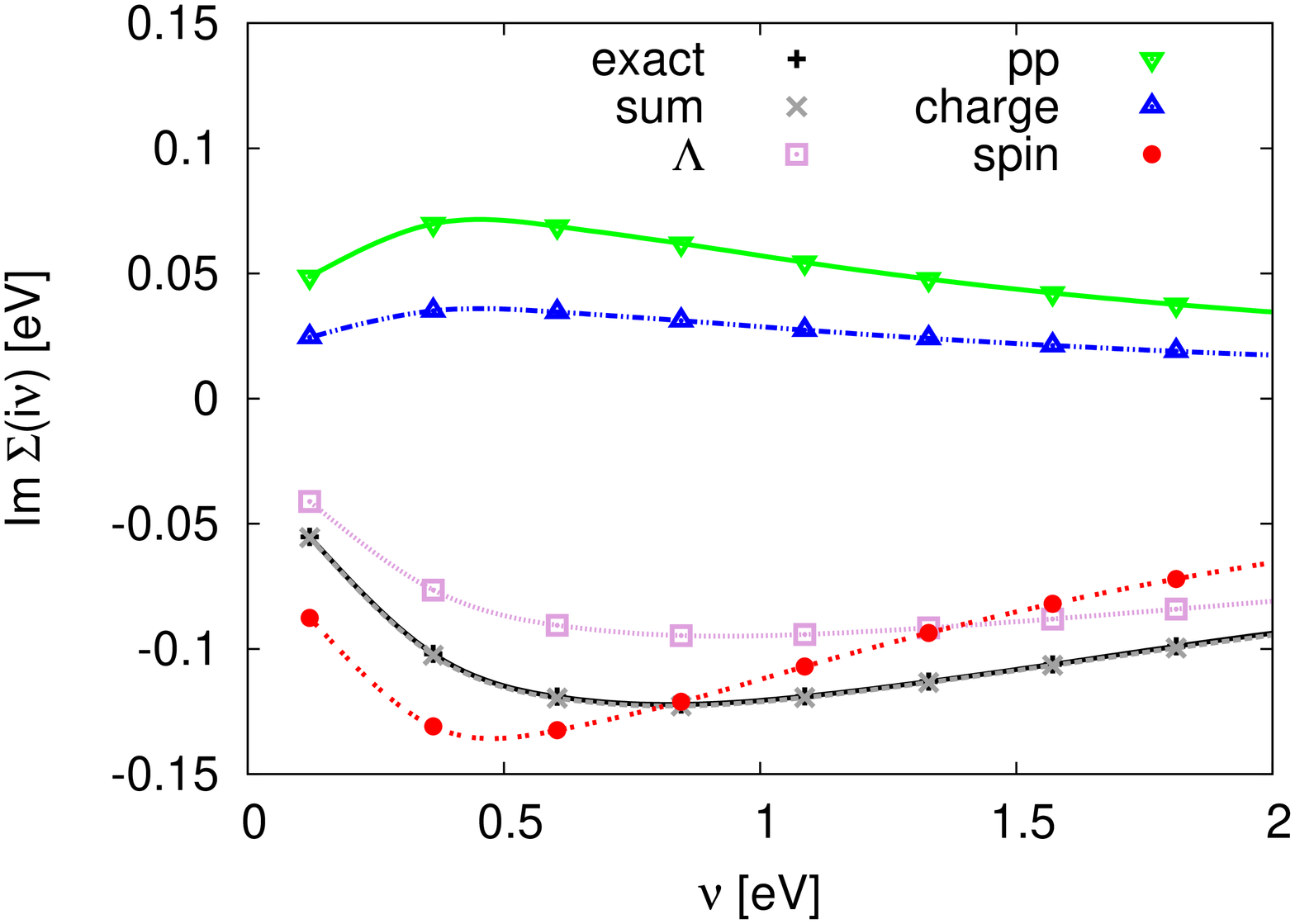}}}}
\caption{(Color online) Parquet decomposition of the DMFT self-energy $\Sigma(\nu)$ of 
the $3d$ Hubbard model at half-filling ($n=1$). The full (black, "exact") and 
dashed (gray, "sum") lines show $\Sigma$ as computed in DMFT, and as the sum 
of the parquet contributions, respectively.  The colored symbols display the 
different contributions to $\Sigma(\nu)$ according to Eq.~(\ref{eq:2.9bis}). 
The parameters of the calculation are: $N_c=1$ (DMFT), 
$t=-\frac{1}{2\sqrt{6}}\simeq -0.204$ eV, $\beta=26$ eV$^{-1}$ with 
two different values of the Hubbard interaction: $U=0.5$ eV (left panel), $U=1$ eV (right panel).}
\label{fig:3.1}
\end{figure*}

In the following section, we will apply this idea to specific cases of interest. In particular, we will test the performance of a parquet decomposition of the self-energy in the case of the three and two-dimensional Hubbard model on a simple cubic/square lattice, whose Hamiltonian reads
\begin{equation}
H =t \sum_{ ij ,\sigma} (c^\dagger_{i \sigma}
c^{\phantom \dagger}_{j \sigma} + c^\dagger_{j \sigma} c^{\phantom \dagger}_{i \sigma})
+ U \sum_{i} n_{i\uparrow} n_{i\downarrow}, 
\label{eq:Hubbard}
\end{equation}
where
$n_{i\sigma}=c^{\dagger}_{i\sigma}c^{\phantom\dagger}_{i\sigma}$,  $t$
the  hopping integral and $U$ is the on-site Coulomb interaction. For
the sake of definiteness, $t=-0.25$ eV for the $2d$ case, and $t=-
\frac{1}{2\sqrt{6}} \simeq -0.204$ eV for the 3d case. This choice ensures that the
standard deviation ($D$) of the non-interacting DOS of the square
and the cubic lattices considered is exactly  the same ($D=1$eV), and
thus allows for a direct comparison of the $U$ values used in the
two-cases, provided they are expressed in units of $D$.

 This Hamiltonian constitutes an important testbed case for applying the idea of a parquet decomposition, since Eq. (\ref{eq:Hubbard}) provides a quintessential representation of a strongly correlated system. Moreover, in the 2d case Eq. (\ref{eq:Hubbard}) is frequently adopted, e.g., to study the still controversial physics of cuprate superconductors.\cite{Dagotto,Scalapino}  In this framework, we note that typical values for $U$ are about $U=8|t|=2$eV, i.e., $U$ is  equal to the non-interacting bandwidth $W=8|t|$. This choice corresponds to a rather strong  correlation regime, as it is clearly seen even in a purely DMFT context.\cite{Toschi2005} In this work, however, we will also consider smaller values  of $U$, of the order of half bandwidth, corresponding to a regime of more moderate correlations.

\section{Parquet decomposition calculations}\label{sec:3}

In this section we study the parquet decomposition of an electron
self-energy computed by DMFT and DCA. In these non-perturbative methods a 
cluster with $N_c$ sites is embedded in a self-consistent electronic bath. 
The calculation of a generalized susceptibility is rather time-consuming when compared against computing only single-particle quantities.  For this reason we restrict our calculations to the tractable values of $N_c=1$ (DMFT), $4$ and $8$ (DCA). The results are 
therefore not fully converged with respect to $N_c$, but, nevertheless, will  
illustrate well the specific points we make in the following sections.
The cluster problem has been solved using both Hirsch-Fye\cite{HirschFye} 
and continuous time (CT)\cite{CT} methods.  

Consistent with the discussion of the previous section, we will use
Eq.~(\ref{eq:2.9}),  illustrated diagrammatically in
Fig.~\ref{fig:2.2}, and Eqs.~(\ref{eq:2.4}), (\ref{eq:2.4bis})
 to express the self-energy in terms of contributions from the different parquet channels. 
As for the latter, in Fig.~\ref{fig:3.0} we show some typical diagrams, 
and their  classifications according to the parquet decomposition. Using the 
definitions in Sec.~\ref{sec:2}, Fig.~\ref{fig:3.0}a and b show 
longitudinal and transverse particle-hole reducible diagrams, 
respectively, and Fig.~\ref{fig:3.0}c shows a particle-particle 
reducible diagram. In fact, the vertex diagram in Fig.~\ref{fig:3.0}a 
contains  contributions to the random phase approximation for the 
longitudinal charge and spin susceptibilities, reducible in spin- and charge-channel. In the same way,
the diagram in Fig.~\ref{fig:3.0}b contains a contribution to 
the transverse spin susceptibility and
Fig.~\ref{fig:3.0}c displays a 
particle-particle ladder diagram.

\begin{figure*}[t!]
{\rotatebox{-90}{\resizebox{6.0cm}{!}{\includegraphics {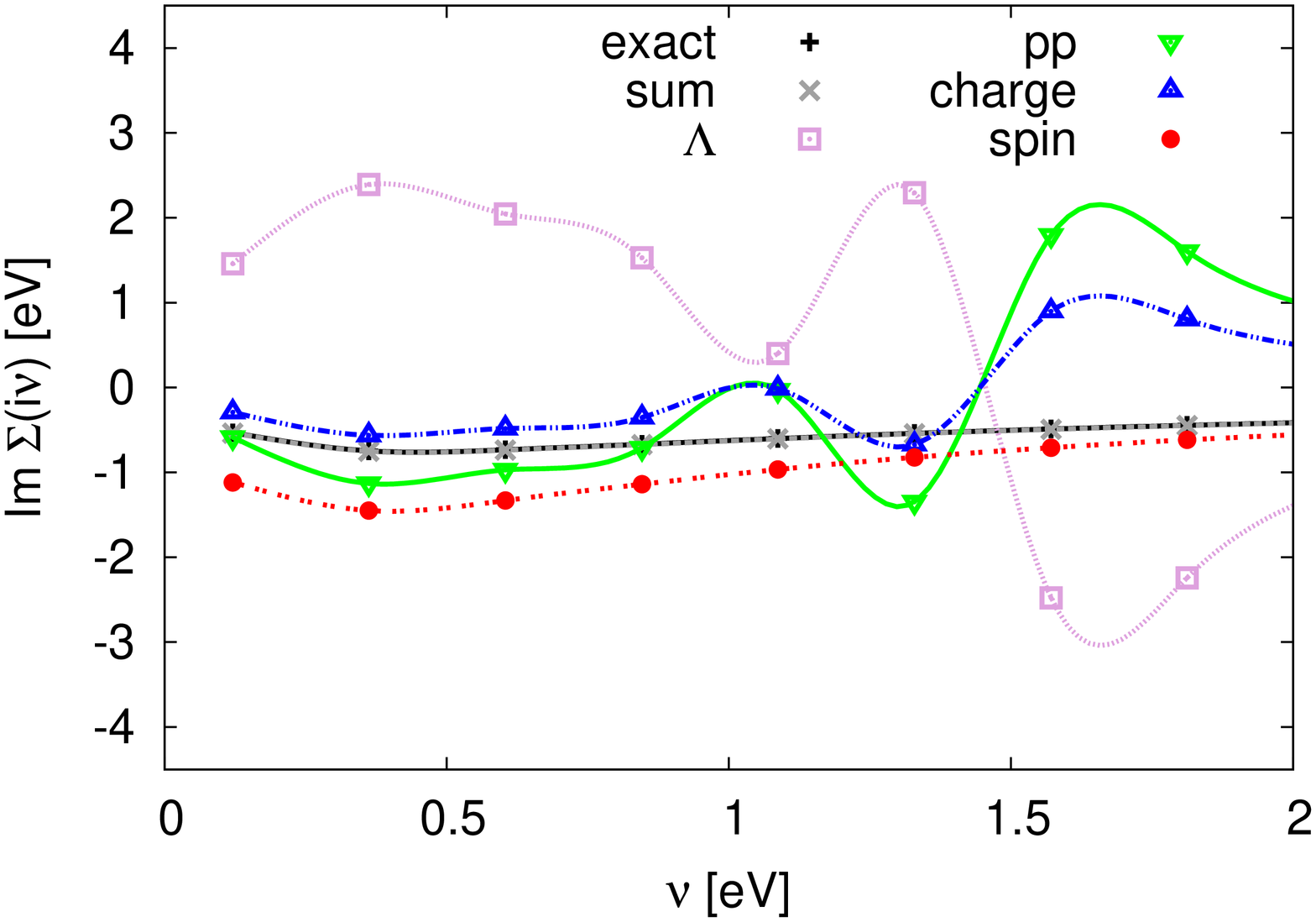}}}}
{\rotatebox{-90}{\resizebox{6.0cm}{!}{\includegraphics {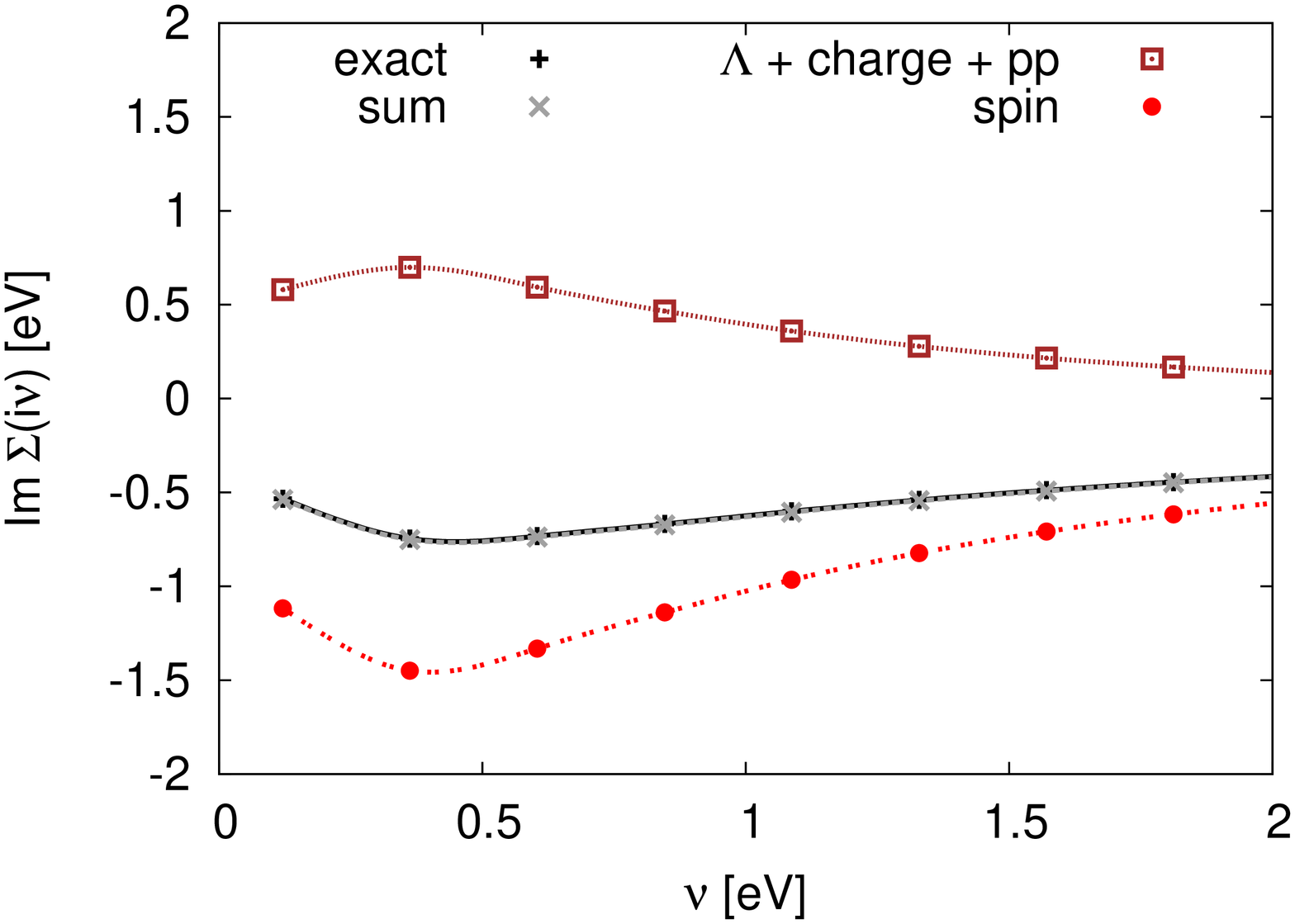}}}}
\caption{(Color online) left panel: Parquet decomposition of the DMFT self-energy $\Sigma(\nu)$ as in Fig.~\ref{fig:3.1}, but with $U=2$ eV. Right panel: Bethe-Salpeter decomposition in the spin channel of the same DMFT self-energy.}
\label{fig:3.2}
\end{figure*}

\subsection{DMFT results}\label{sec:3.1}

We start by applying the parquet decomposition to the easier case of 
the DMFT self-energy.
In particular, we will focus on one of the most studied cases in DMFT, 
the half-filled Hubbard model in $3d$, where DMFT describes a 
Mott-Hubbard metal-insulator transition at a finite $U=U_{MIT}$. 
The specific parameters in Eq.~(\ref{eq:Hubbard}) have been chosen 
in this case as follows: $n=1$ (half-filling) and $\beta=26$ eV$^{-1}$.
The results of the parquet decomposition of the DMFT self-energy are 
shown in Fig.~\ref{fig:3.1} in the weak-to-intermediate coupling 
regime $U\ll{U_{MIT}}\sim 3$ eV. The plots show the imaginary part 
of the DMFT self-energy (solid black line) as a function of 
the Matsubara frequencies i$\nu$  and for two different values 
of $U$ (we recall that $\Sigma$ does not depend on momentum in DMFT, 
and that in a particle-hole symmetric case, as the one we consider here, 
it does not have any real part beyond the constant Hartree term). 

By computing the DMFT generalized local ($N_c=1$) susceptibility 
of the associated impurity problem, and proceeding as described 
in the previous section, we could actually decompose Im $\Sigma(i\nu)$ 
into the four contributions from terms in Eq.~(\ref{eq:2.9bis}), depicted by different 
colors/symbols in the plots. Before analyzing their specific behaviors, 
we note that their sum (gray dashed line) {\sl does} reproduce 
precisely the value of Im $\Sigma$ directly computed 
in the DMFT algorithm. Since all the four terms of Eq.~\ref{eq:2.9bis} 
are calculated independently from the parquet-decomposed equation 
of motion, this result represents indeed a stringent test of the 
numerical stability and the algorithmic correctness of our parquet 
decomposition procedure. Given 
the number of steps involved in the algorithm, 
illustrated in the previous section, the fulfillment of such a 
self-consistency test is particularly significant, and, indeed, 
it has been verified for all the parquet decomposition 
calculations presented in this work.      
  
By considering the most weak-coupling data first ($U=0.5$eV, left
panel of Fig.~\ref{fig:3.1}), we note that the 2PI contribution
($\Sigma_\Lambda$ in Eq.~(\ref{eq:2.9bis}), plum-colored open squares
in the Figure) lies almost on top of the ``exact'' DMFT self-energy. 
At weak-coupling this is not particularly surprising, because
$\Lambda_{\uparrow\downarrow} \simeq U + O(U^4)$, while all the 2PR
contributions are at least $O(U^2)$.
Hence, when the 2PI vertex is inserted into the equation of motion,
$\tilde{\Sigma}_\Lambda$ simply reduces to the usual second-order
perturbative diagram. In this situation (i.e., Im $\Sigma(i\nu) \simeq
\tilde{\Sigma}_\Lambda$), it is interesting to observe that the
other sub-leading contributions (spin, particle-particle scattering and charge channel) 
are not fully negligible. Rather, they almost exactly compensate each other: 
the extra increase of the scattering rate [i.e.: -Im $\Sigma(i\nu \rightarrow 0)$] due to the spin-channel 
is compensated (or ``screened'') almost perfectly by the charge- and the particle-particle channel.

Not surprisingly, the validity of this cancellation is gradually lost by
increasing $U$. At $U=1.0$ (right panel of Fig.~\ref{fig:3.1}), which
is still much lower than $U_{MIT}$, one observes that the 2PI
contribution no longer provides so accurate values for Im
$\Sigma(\nu)$. At the same time, the contributions of all scattering
channels increase: the low-frequency behavior of the spin channel now
would provide -taken on its own- a scattering rate even larger than
the true one of DMFT. Consistently, a correspondingly larger
compensation of the charge and the particle-scattering channels
contribution is observed. At higher frequency, these changes
w.r.t. the previous case are mitigated, matching the intrinsic
perturbative nature of the high-frequency/high-T expansions\cite{Kunes2011,Georgthesis,Stefanthesis}. 

The situation described above, which suggests an important role of spin fluctuations, partially
screened by charge and particle-particle scattering processes, displays
important changes at intermediate-to-strong coupling $U$. This is well
exemplified by the data reported in Fig.~\ref{fig:3.2} (left
panel). Despite the DMFT self-energy still displays a low-frequency
metallic bending ($U=2.0$ is on the metallic side of the DMFT MIT), in
the low-frequency region 
one observes the appearance of a huge oscillatory behavior in the parquet
decomposition of $\Sigma$: All  contributions to Im $\Sigma$, {\sl
  but} the spin term (s. below), are
way larger than the self-energy itself and fluctuate so strongly in
frequency, that several changes of sign are observed. This makes it
obviously very hard to define any kind of hierarchy for the impact of the
corresponding scattering channels on the final self-energy result. 

Hence, at these intermediate-to-strong values of $U$ the parquet
decomposition procedure appears to be no longer able to fully disentangle the
physics underlying a given (here: DMFT) self-energy. 
At the same time, we should stress that the strong oscillations
visible in the parquet decomposition of Fig.~\ref{fig:3.2} can {\sl
  not} be ascribed to numerical accuracy issues. In fact, one
observes, that, also in this problematic case, the self-consistency test
works as well as for the other data sets:  the total sum of such oscillating
contributions, still reproduces the Im $\Sigma(\nu)$ from DMFT  in the
whole frequency range considered.  The reason of such behavior
has to be traced back, instead, to the
divergencies of the 2PI vertices recently reported in DMFT work.\cite{divergence,Georgthesis,Schaeferthesis,Janis2014,Kozik2015} 
While the relation with such divergencies will be extensively
discussed in Sec. IV, it is worth stressing already here, that there is only one
contribution to $\Sigma(\nu)$, which never displays wild oscillation,
even for intermediate-to-strong $U$: the spin channel. 
This means that even when the parquet decomposition displays a strong
oscillatory behavior, a Bethe-Salpeter
decomposition in this specific (spin) channel will always remain 
well-behaved and meaningful. This is 
explicitly shown in Fig.~\ref{fig:3.2} (right panel), where all the
contributions to $\Sigma(\nu)$, but $\tilde{\Sigma}_{sp}$, (i.e., formally: all the 
contributions 2PI in the spin channel) are summed together: Here no
oscillation is visible. 
The results of such Bethe-Salpeter decomposition of $\Sigma(\nu)$ in
the spin  channel suggests then again an interpretation of a physics dominated
by this scattering channel, though -this time- in the non-perturbative
regime: Strong (local) spin fluctuations, originated
by the progressive formation of localized magnetic moments, are
responsible for the major part of the electronic self-energy and scattering rate.  
Their effect is, as before, partly reduced, or screened, by the
scattering processes in the other channels (opposite sign contribution
to Im $\Sigma$).  Differently as before, however, the specific role of
the ``secondary'' channels can no longer be disentangled via our
parquet decomposition.

\subsection{DCA results}\label{sec:3.2}

\begin{figure*}
{\rotatebox{-90}{\resizebox{6.0cm}{!}{\includegraphics {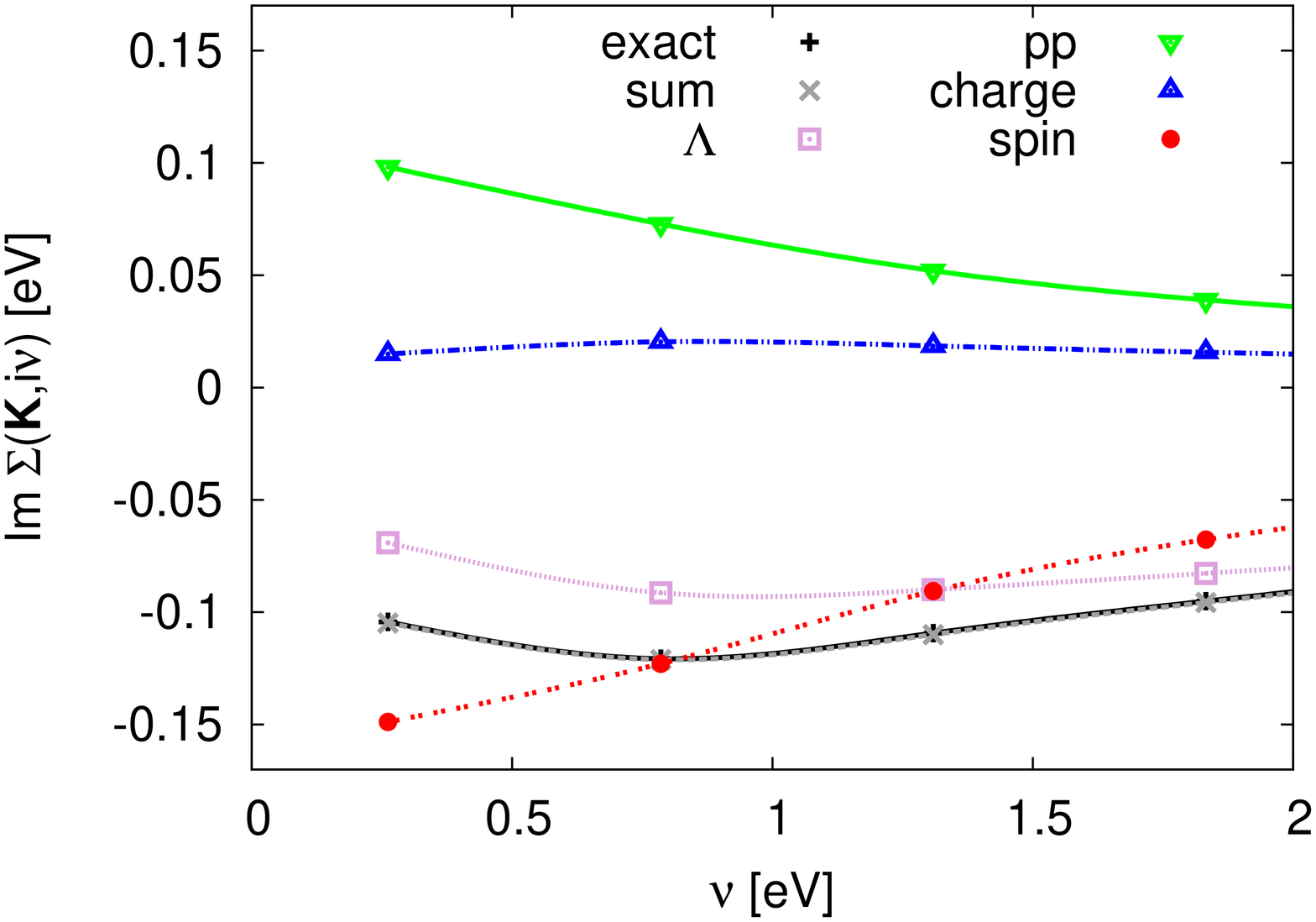}}}}
{\rotatebox{-90}{\resizebox{6.0cm}{!}{\includegraphics {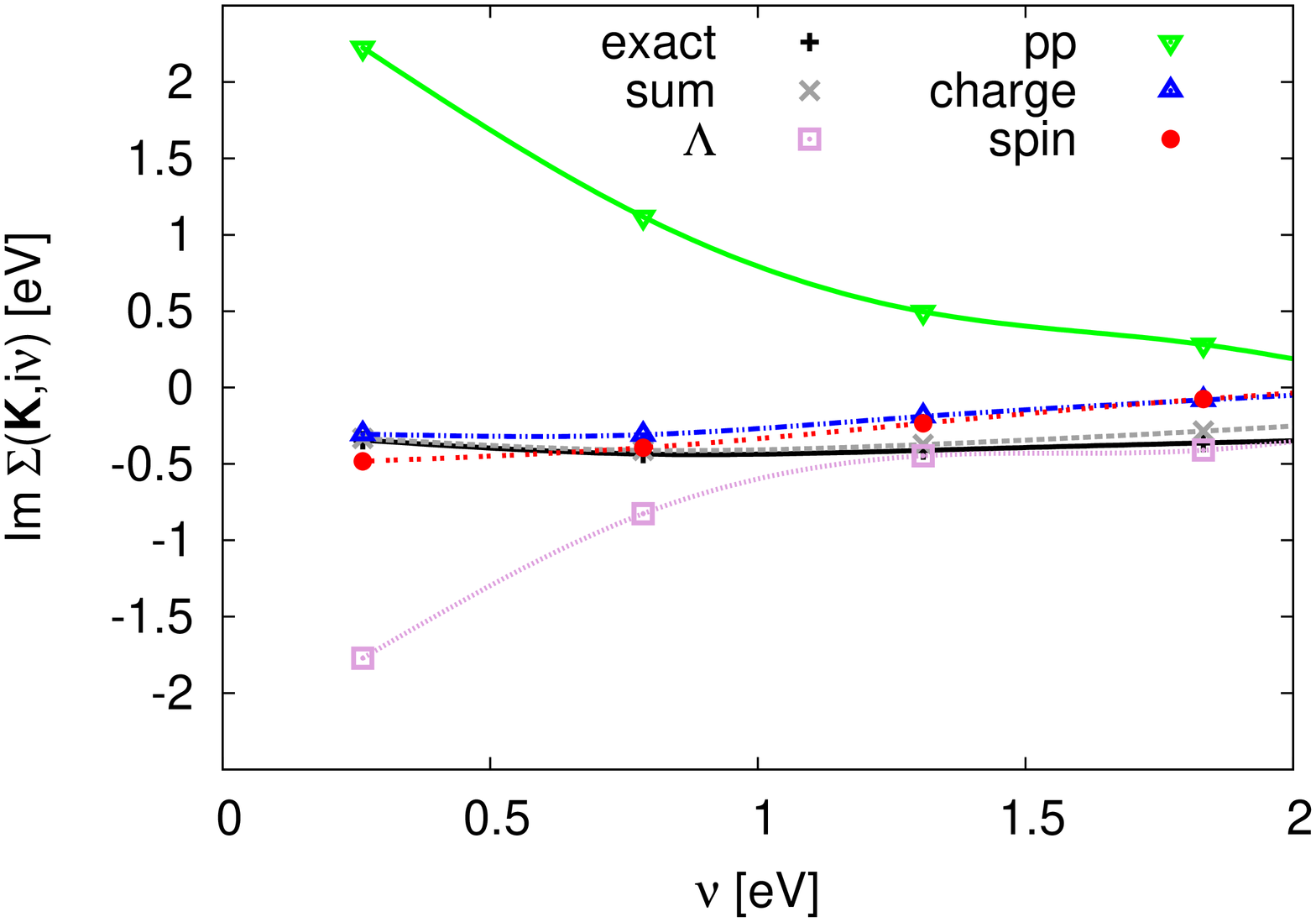}}}}
\caption{(Color online) Parquet decomposition of the DCA self-energy
  $\Sigma[{\bf K}=(\pi,0),\nu]$. The same convention of Fig.~\ref{fig:3.1} is adopted. 
The parameters of the calculations are $N_c=8$, $t=-0.25$ eV, $\beta=12$ eV$^{-1}$ and 
the filling is $n=0.85$ with  two different values of the Hubbard interaction: $U=1.0$ eV (left panel), 
$U=2.0$ eV (right panel)\label{fig:3.3}}
\end{figure*}

In this subsection, we discuss the numerical results for the parquet decomposition of  self-energy data computed in DCA.
Different from DMFT, the DCA self-energy provides a more accurate description of finite dimensional systems,
as it is also explicitly dependent on the momenta of the 
discretized Brillouin zone (i.e., a cluster of $N_c$ patches in momentum space) of the DCA. 
We will present here parquet decomposition results for the self-energy of the two-dimensional Hubbard model with hopping parameter $t=-0.25$ for different values of 
the density $n$ and of the interaction $U$. In particular, we will mostly focus on the self-energy at the so called anti-nodal point, ${\bf K}=(\pi,0)$, 
because it usually displays the strongest correlation effects for this model and also because the vector  ${\bf K}=(\pi,0)$ is always present in both clusters we used ($N_c=4, 8$) in our DCA calculation. We note, however, that the results of the parquet decomposition
for the other relevant momenta of this system, i.e. the nodal one ${\bf K}=(\pi/2,\pi/2)$, (for $N_c=8$ where it is available), are qualitatively 
similar. 

As for the DMFT case, we start by considering a couple of significant cases at fixed density (here $n=0.85$, corresponding to 
the typical $15$\% of hole doping of the optimally doped high-T$_c$ cuprates), and perform the parquet decomposition
for different $U$. In the left panel of Fig.~\ref{fig:3.3} we show the calculations performed at a moderate $U=4|t|=1$eV (interaction equal to the semibandwidth). As one sees the results are qualitatively similar to the DMFT one at intermediate coupling (right panel of Fig.~\ref{fig:3.1}), which one could indeed interpret in terms of predominant spin-scattering processes, partially screened by the other channels. However, also in DCA, extracting such information from the parquet decomposition becomes rather problematic for larger values of $U$. At $U= 8|t| =2$ eV (interaction equal to the bandwidth: Fig.~\ref{fig:3.3} right panel), the parquet decomposition appears dominated by contributions from the 2PI and the $pp$ channel: These become an order of magnitude larger than the spin-channel contribution and of the total DCA self-energy. This finding, in turn, indicates the occurrence of large cancellation effects in the parquet-decomposed basis, making quite hard any further physical interpretation.  

\begin{figure*}[t]
{\rotatebox{-90}{\resizebox{6.0cm}{!}{\includegraphics {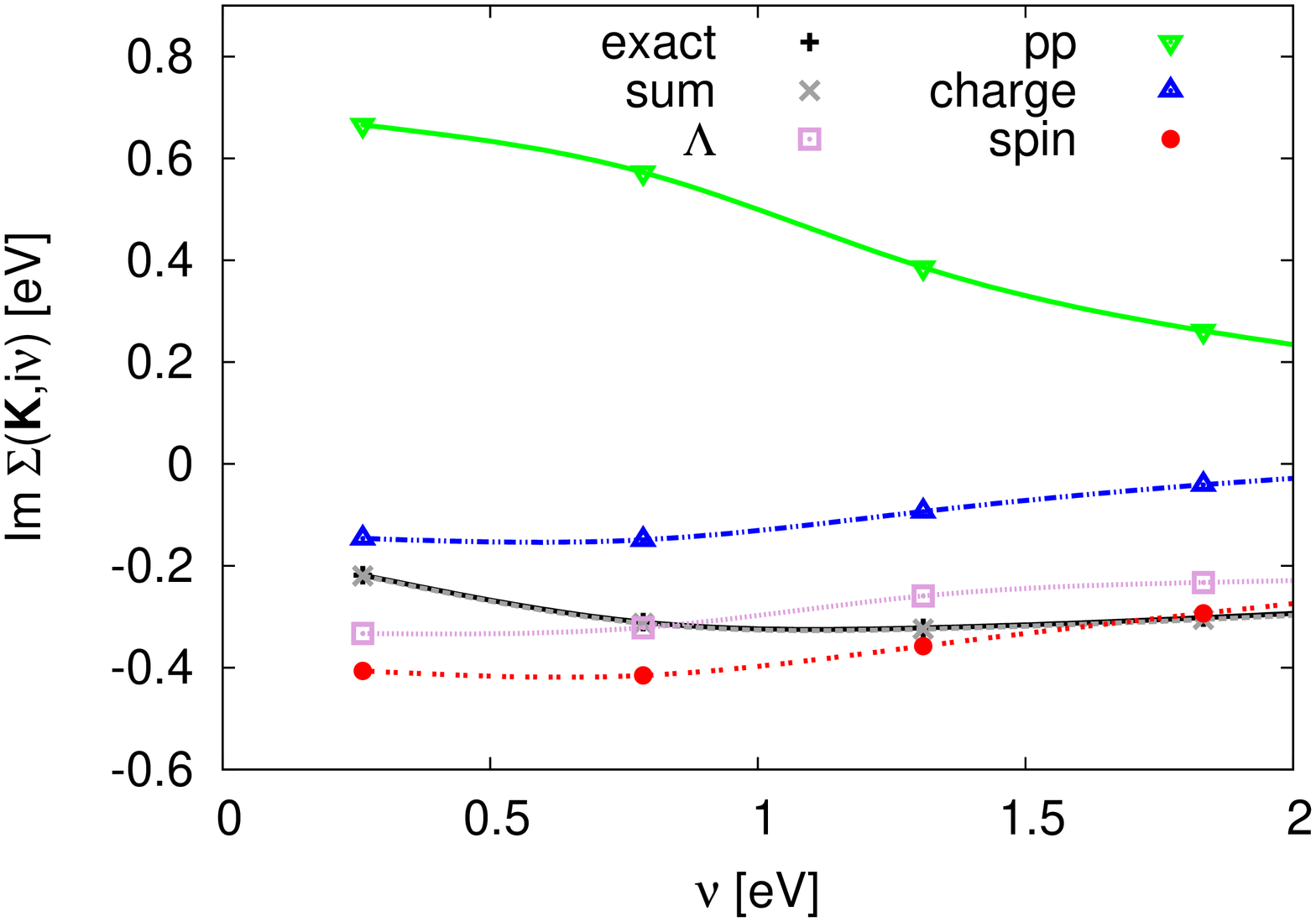}}}}
{\rotatebox{-90}{\resizebox{6.0cm}{!}{\includegraphics {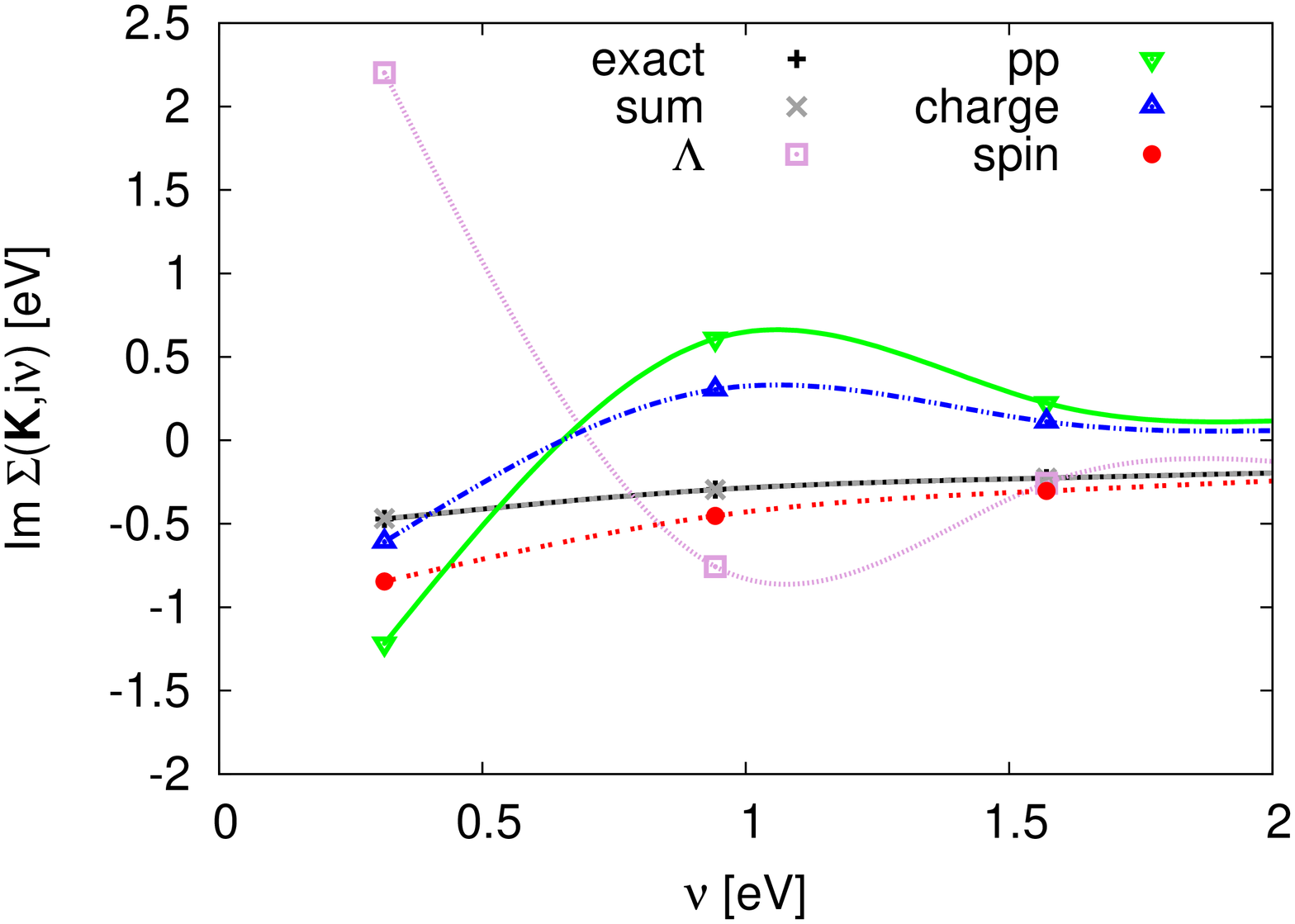}}}}
\caption{Parquet decomposition of the DCA self-energy $\Sigma[{\bf K}=(\pi,0),i\nu]$ ($N_c=8)$ at different dopings. Left panel: high hole doped case ($n=0.75$) for the same interaction/temperature values as in right panel of Fig.~\ref{fig:3.3} ($U=2$ eV and $\beta=12$ eV$^{-1}$). Right panel: undoped case ($n=1$), at  intermediate-to-strong coupling ($U=1.4$ eV and $\beta=10$ eV$^{-1}$, calculated for a $N_c=4$ DCA cluster.
\label{fig:3.4}}
\end{figure*}

It is also instructive to look at the effect of a change in the
level of hole-doping on the parquet decomposition calculations. This is
done in Fig.~\ref{fig:3.4}: In the left panel of the figure results
for the highly doped case $n=0.75$ ($25$\% hole doping) are
shown. Despite the large value of the interaction $U=2$ eV, this
parquet decomposition looks qualitatively similar to the one at
moderate coupling of the less doped case  Fig.~\ref{fig:3.3} (left
panel). Conversely, at half-filling ($n=1$, right panel of
Fig.~\ref{fig:3.4}), although we chose a lower value of $U=1.4$ eV,
the parquet decomposition displays the very same large oscillations
among 
different channel contributions observed in the DMFT data (Fig.~\ref{fig:3.2}, left panel). 
Hence, our parquet decomposition procedure applied to the DCA results
allows us to extend the considerations drawn from the DMFT analysis of
the previous section: For a large enough value of $U$ and moderate or
no doping, the parquet decomposition of the self-energy becomes rather
problematic, as some channel contributions (supposed to be secondary)
become abruptly quite large, or even strongly oscillating, with large
cancellation between different terms.  The inclusion of non-local
correlations within the DCA allows us to demonstrate that this is {\sl
  not} a special aspect of the peculiar, 
purely local, DMFT physics, but it survives also in presence
of non-local correlations. Actually, as we will discuss in the next
sections, the non-local correlations do {\sl favor} the occurrence of
singularities in the parquet decomposition, which is observed for DCA
in a correspondingly larger parameter region (at lower $U$ and
hole-doping) than in DMFT.

\begin{figure*}
{\rotatebox{-90}{\resizebox{6.0cm}{!}{\includegraphics {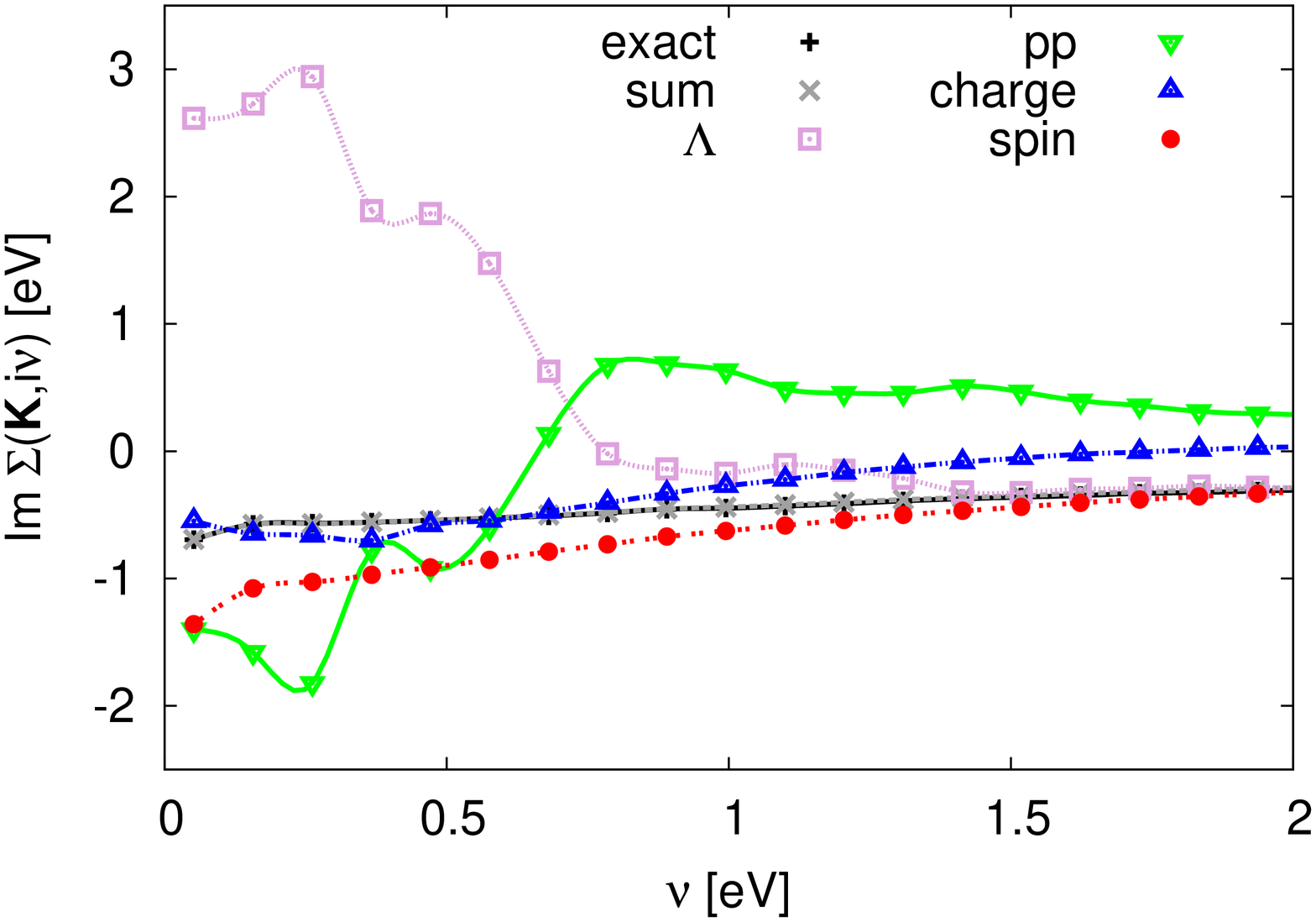}}}}
{\rotatebox{-90}{\resizebox{6.0cm}{!}{\includegraphics {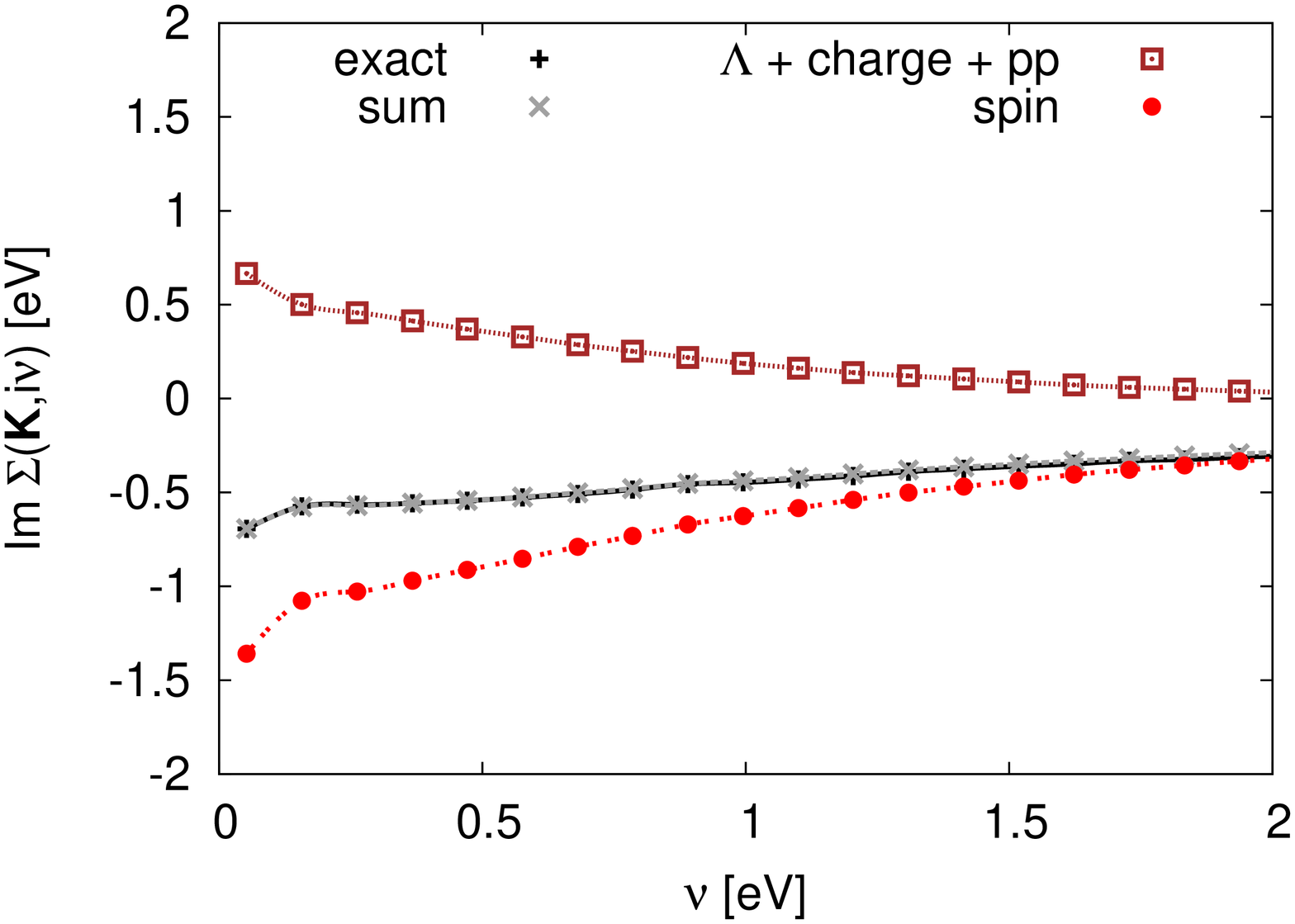}}}}
{\rotatebox{-90}{\resizebox{6.0cm}{!}{\includegraphics {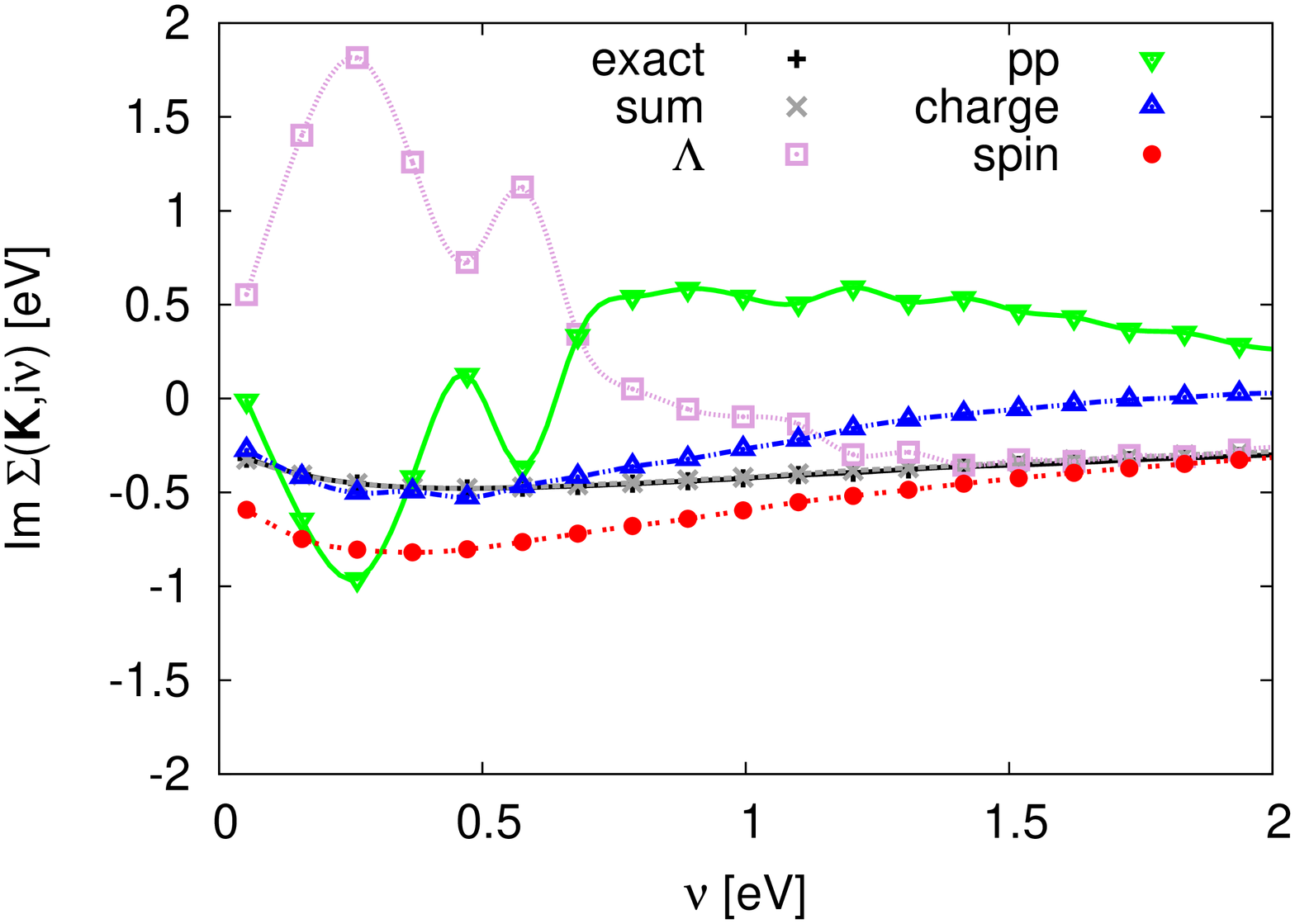}}}}
{\rotatebox{-90}{\resizebox{6.0cm}{!}{\includegraphics {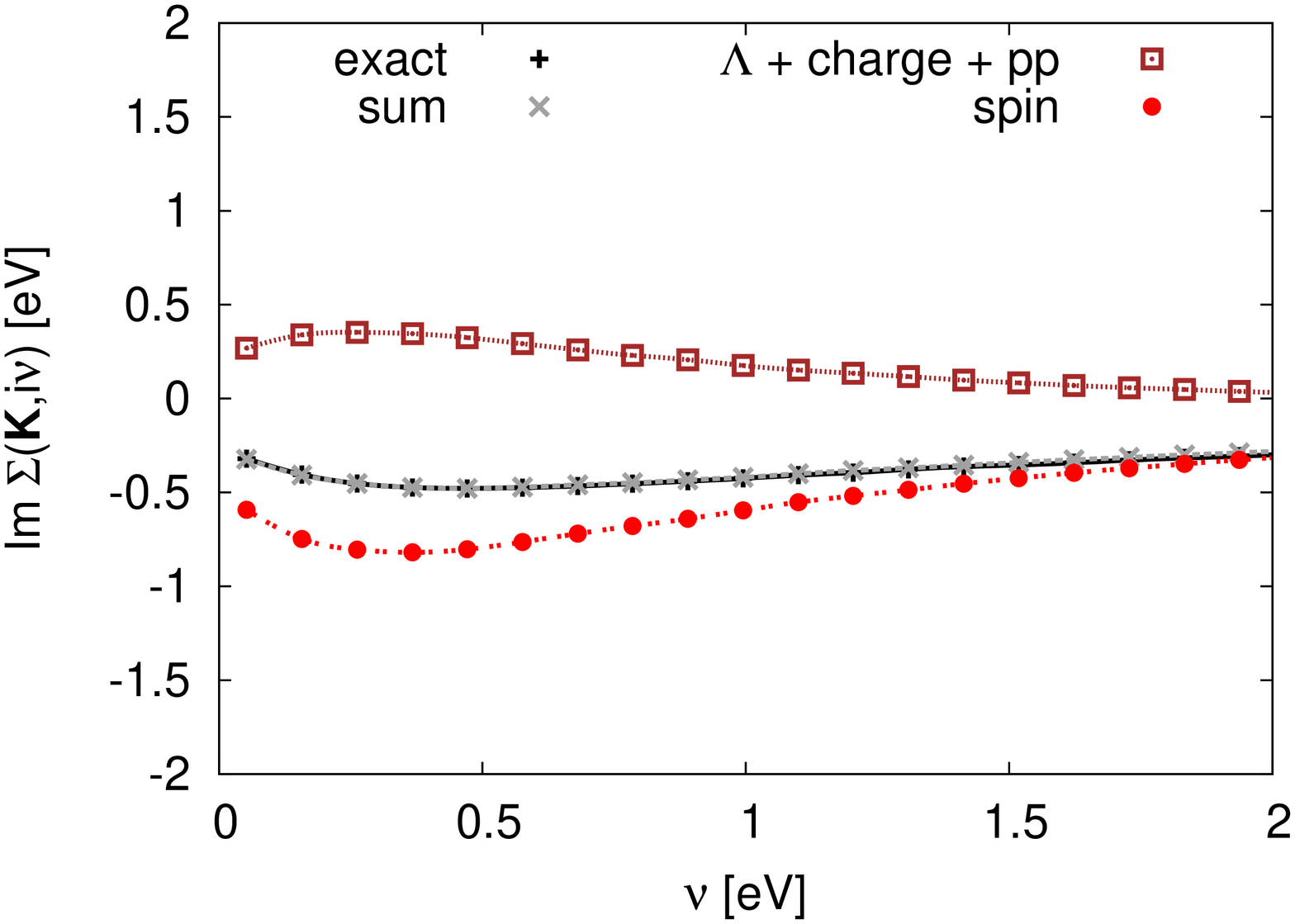}}}}
\caption{Parquet decomposition of the DCA self-energy $\Sigma[{\bf K},i\nu]$ with $N_c=8$ for 
the low-$T$, underdoped case $n=0.94$ with $U=1.75$ eV and $\beta=60$ eV$^{-1}$ (see text). Left upper panel: parquet decomposition
for the antinodal DCA self-energy [${\bf K}=(\pi,0)$]; right upper panel: Bethe-Salpeter decomposition of the
antinodal DCA self-energy. Left lower panel: parquet decomposition of the nodal[${\bf K}=(\frac{\pi}{2},\frac{\pi}{2})$] 
DCA self-energy. Right lower panel: Bethe-Salpeter decomposition of the nodal DCA self-energy.
\label{fig:3.5}}
\end{figure*}

In this perspective, it is interesting to investigate, whether the
singularities in the parquet decomposition, with their intrinsically non-perturbative nature, already occur in a parameter region where the DCA self-energy displays a  strong momentum differentiation, with pseudogap features. As discussed in Ref.~\onlinecite{FluctDiag}, such a case is achieved in a $N_c=8$ DCA calculation for, e.g.,: $n=0.94$ ($6$\% hole doping), $U=1.75$ eV, $\beta=60$ eV$^-1$ (with the additional inclusion of a realistic next-to-nearest hopping term $t'=0.0375$ eV).
In the left panels of the Fig.~\ref{fig:3.5} the DCA self-energy for
the anti-nodal and the nodal momentum is shown, together with its
corresponding parquet decomposition. We note, as it was also stated in
Ref.~\onlinecite{FluctDiag}, that the positive (i.e., non Fermi-liquid) slope of Im$\Sigma({\bf
  K},i\nu)$ in the lowest frequency region for $\mathbf{K}=(\pi,0)$ indicates a pseudogap
spectral weight suppression at the antinode. The parquet decomposition
of the two self-energies is, however, very similar: The strong
oscillations of the various channels clearly demonstrate that in the
parameter region where a pseudogap behavior is found in DCA, the
parquet decomposition displays already strong oscillations.
It is also interesting to notice that, similarly as we discussed in the
previous section, also in this case, the spin channel contribution of
the parquet decomposition is the only one displaying a well-behaved
shape, with values of the order of the self-energy and no frequency
oscillations. Consequently, also for the DCA self-energy in the pseudogap
regime, a Bethe-Salpeter decomposition in the spin-channel of the
self-energy remains valid (see right panel of Fig.~\ref{fig:3.5}). As
discussed in the previous section, this might be interpreted as an
hallmark of the predominance of the spin-scattering processes in a
non-perturbative regime, where a well-behaved parquet decomposition is
no longer possible. In this perspective, the physical interpretation
would match very well the conclusions derived
about the origin for the pseudogap self-energy of DCA by means of the
recently introduced fluctuation diagnostics method\cite{FluctDiag}. At
present, hence, the post processing of a given numerical
self-energy provided by the fluctuation diagnostics procedure appear
the most performant, because -differently from the parquet decomposition- it remains 
applicable, without any change, also to non-perturbative cases.  

After discussing our parquet decomposition calculations, their proposed physical interpretation, and their limitation in applicability, it is natural to wonder, where such limitations arise from. This analysis is, in fact, very important also {\sl beyond} the calculations presented in this work, because the parquet equations represent the base-camp of several novel quantum many body schemes aiming at the the description of strongly correlated electron beyond the perturbative regime.  

As we anticipated before, the reason for the occurrence of strong
low-frequency oscillations in the parquet decomposition can be traced to the {\sl divergence} of the 2PI irreducible vertex functions observed by increasing $U$,\cite{divergence} or -equivalently- to the occurrence of {\sl singularities} in the  generalized $ph$ charge ($\chi_{ch}$) and $pp$ ($\uparrow\downarrow$ and/or singlet) ($\chi_{pp}$) susceptibilities. 
The investigation of the exact relation between the peculiar behavior 
of the parquet decomposition by increasing $U$ and the singularities of the corresponding generalized susceptibility matrix will be explicitly addressed below.

\section{Singularities of  generalized susceptibilities}\label{sec:4}

In this section, we aim at clarifying {\sl why} some contributions of the parquet decomposition
start displaying singularities and strong oscillatory behaviors upon increasing $U$. 
From a general perspective, since the singularities observed in
the previous section {\sl always} affect $\tilde{\Sigma}_\Lambda$,
 the contribution stemming from the 2PI vertex, 
a clear relation must exist with corresponding divergencies
of the 2PI vertices. In fact, the occurrence of
divergences in the 2PI vertices of the Hubbard and Falicov-Kimball
model has been recently demonstrated by means of analytic and DMFT
calculations.\cite{divergence, Janis2014} In particular, we recall
that such singularities show up simultaneously in the fully 2PI vertex
$\Lambda$ as well as in the irreducible vertices $\Gamma_r$ in the
charge ($r=ch$) and particle-particle
channel($r=pp,\uparrow\downarrow$), while the full vertex $F$ and the
self-energy remain always well-behaved. Evidently, this perfectly matches the problematic channels of our parquet decompositions.

As discussed in Ref.~[\onlinecite{divergence}], a
divergence of a $\Gamma_r$ must be associated to a non-invertibility
of its corresponding Bethe-Salpeter equation and, hence, according to
Eq.~(\ref{eq:2.6}), to the occurrence of singular ($= 0$) eigenvalue in the
generalized susceptibility matrix $\chi_{r}(k;k';q)$.  
In fact, when an eigenvalue goes through zero, the irreducible vertex functions
change qualitatively. In particular, one observes that second order 
perturbation theory breaks down, failing to reproduce even the sign 
of the vertex-functions at low frequencies. In this sense the system
is then in the truly strong-coupling limit. 
In Ref.~\onlinecite{divergence}, $\chi_{ch}(k;k';q)$ was computed in DMFT
($N_c=1$), treating $\chi_{ch}$ as a matrix in $k$ and $k'$ for fixed $q$. For the case when the 
frequency transfer $\omega$ is zero, we showed that the lowest 
eigenvalue of this matrix becomes negative as $U$ is increased. 
A similar behavior was found for $\chi_{pp}$ in the $\uparrow\downarrow$ sector 
(or in the singlet channel) for a somewhat larger $U$.

In the following, we will analyze in more details such divergencies, by extending
the previous DMFT ($N_c=1$) results\cite{divergence} to DCA ($N_c=4$), 
and by investigating in details how singularities develop in the
generalized susceptibility matrices and how they affect the parquet decompositions of the self-energy.

\subsection{$N_c=1$ case}\label{sec:4.1}

For the sake of clarity we start by analyzing the generalized charge
susceptibility in the $N_c=1$ (DMFT) case, focusing on the most-correlated case of half-filling. 
In particular we will mainly study the most singular case of $\omega=0$.  In
fact, $\omega=0$ represents the largest contribution to
the parquet decomposition for the values of $T$ studied here, and, thus, its behavior 
is particularly significant. The case $\omega\ne 0$, nonetheless, will
be also discussed briefly afterwards. 

For a very small value of $U$,  where no problem in the parquet
decomposition is observed, we can approximate the generalized charge susceptibility
with the non interacting one, i.e., with a product of two Green's functions 
$\chi_{ch}(\nu,\nu', \omega=0) \simeq \chi^0(\nu,\nu',\omega=0)\delta_{\nu, \nu'}$. 
In addition we can use noninteracting Green's functions. The corresponding 
diagonal elements are given by   
\begin{equation}\label{eq:5.1}
\chi_{ch}(\nu;\nu;\omega=0) \simeq -{\beta\over N_k^2}\sum_{{\bf k},{\bf k}'} 
{1 \over (i\nu+\mu-\varepsilon_{\bf k} )(i\nu+\mu-\varepsilon_{{\bf k}'})},
\end{equation}
where $N_k$ is the number of ${\bf k}$-points and $\varepsilon_{\bf k}$ 
is the corresponding single-particle energy eigenvalue.  Off-diagonal elements will obviously
appear at finite $U$, remaining however much smaller than the diagonal ones
in the perturbative regime. If we now consider the limit of very large
$\nu$, the diagonal elements behave as $\beta/\nu^2$ and, hence,
also become very small. In the numerical calculations, we 
limit the range of $|\nu|$ to some maximum value $\nu_{\rm max}$. Hence, in the perturbative regime, the lowest 
eigenvalue of $\chi_{ch}$ will correspond roughly to the value of the diagonal element 
for $\nu=\nu_{\rm max}$, and its eigenvector will have 
weight for $\nu=\pm \nu_{\rm max}$. 

 As $U$ is increased, however, the off-diagonal elements $\nu'\ne \nu$ become gradually more important until, at a
certain point,  (e.g., at $U = \bar{U} \sim 1.27$ eV for the temperature we considered)  this 
picture changes radically: The off-diagonal component of $\chi_{ch}(\nu,
\nu', \omega=0)$  for small frequencies become comparable or larger than the corresponding
diagonal ones. As a consequence (see appendix \ref{app:2by2}), the lowest eigenvalue of $\chi_{ch}(\nu,
\nu', \omega=0)$ crosses zero and, for large interaction, a negative eigenvalue appears. 
In contrast to the small $U$ case, the corresponding vector has most of its 
weight for $\nu=\pm \pi/\beta$: For these parameters the total weight
of two elements for $\nu=\pm \pi/\beta$ is about $0.85$.
This indicates that a crossing of energy levels has
occurred between a lowest eigenvector having most of the weight at large
frequencies to the one having most of the 
weight for small frequencies.\cite{notelowT}

\begin{figure}
 {\rotatebox{-90}{\resizebox{5.7cm}{!}{\includegraphics {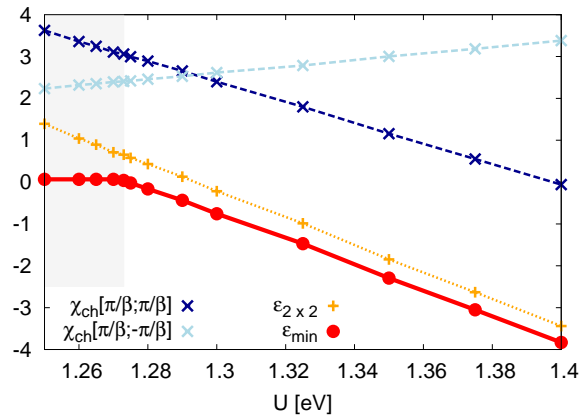}}}}
\caption{\label{fig:5.1} Plot of the lowest eigenvalue of 
  $\chi_{ch}(\nu, \nu',\omega=0)$ as a function of $U$. The parameters
  of the calculation are  $t=-0.25$ eV, $\beta=10$ eV$^{-1}$, $n=1$, and
  $N_c=1$ (DMFT).  The gray
  shadowed area marks the (perturbative) parameter region ($U <
  \bar{U}$), where the level-crossing of the lowest eigenvalue  has
  not yet occurred (see text). The numerical values of
  $\varepsilon_{min}$ are then compared with the approximation in Eq.~(\ref{eq:5.2}).
Finally, the corresponding diagonal and off-diagonal elements of $\chi_{ch}$ 
for $\nu=\pm \pi/\beta$ are also plotted.}
\end{figure}

In this situation, the most significant piece of information can
be extracted by restricting the analysis to the matrix elements 
for $\pm \pi/\beta$, i.e., to a $2\times 2$ matrix in frequency space. Then the lowest 
eigenvalue of $\chi_{ch}$ is
\begin{equation}\label{eq:5.2}
\varepsilon_{\rm 2\times 2}=\chi_{ch}({\pi \over \beta};{\pi \over \beta})-
|\chi_{ch}({\pi\over \beta};-{\pi \over \beta})|.
\end{equation}
This approximation is compared with the exact eigenvalue in Fig.~\ref{fig:5.1}. 
It provides a good approximation after the level crossing has occurred, i.e.,
where the lowest eigenvalue of $\chi_{ch}$ has become negative.
Fig.~\ref{fig:5.1} also shows the elements $\chi_{ch}(\pi/\beta;\pi/\beta)$ 
and $\chi_{ch}(\pi/\beta;-\pi/\beta)$.  The diagonal element decreases and 
the off-diagonal element increases as $U$ is increased. Approximately 
as they cross, the lowest eigenvalue goes negative (the minor deviation of
$\simeq 0.02$U  reflects the corresponding small difference between
$\varepsilon_{\rm 2\times 2}$ and $\varepsilon_{\rm min}$).

As the lowest eigenvalue $\varepsilon_{\rm min}$ of
$\chi_{ch}(\nu,\nu', \omega=0)$ goes through zero, $\chi_{ch}^{-1}$
becomes infinite. For the cases we have studied, the diagonal and 
off-diagonal matrix elements of the $2\times 2$ matrix have the same sign 
when this happens. Consequently, for the corresponding (singular)
eigenvector, the elements for $\nu=\pm \pi/\beta$ then have opposite signs.
It then follows from Eq.~(\ref{eq:5.3}) that the diagonal and off-diagonal 
parts, $\chi_{ch}^{-1}(\pi/\beta;\pi/\beta)$ and $\chi_{ch}^{-1}(\pi/\beta;-\pi/\beta)$ 
also have opposite signs. Inserting $\chi_{ch}^{-1}$ in the expression
needed for the parquet decomposition, we then find that a 
cancellation of the singular contributions does occur (see Appendix \ref{app:2by2}).  

However, as one can easily
infer from the right side of Fig.~\ref{fig:5.1}), at larger values of
$U$, the sign of the diagonal matrix element changes, and then the
(now non-singular) contributions add constructively. Hence, when a
second eigenvalue of $\chi_{ch}(\nu,\nu', \omega=0)$ crosses zero,
no cancellation will occur and the singularity will show up in the
corresponding terms of the parquet decomposition of $\Sigma(i\nu)$.

In particular, from Ref.~[\onlinecite{divergence}] we know that a second
divergence takes place at a slightly larger value of $U$ than the
range of  Fig.~\ref{fig:5.1}. In particular, for $U>1.6$ eV, a second
 eigenvalue of $\chi_{ch}(\nu,\nu', \omega=0)$ vanishes,
 simultaneously with the first one of $\chi_{pp,\uparrow\downarrow}(\nu,\nu', \omega=0)$ 
As discussed above, now, the sign of the matrix elements is 
such that the singular contributions to the parquet decomposition 
{\sl no longer} cancel. Then the parquet decomposition, in its
corresponding counterparts ($\tilde{\Sigma}_{ch}$,
$\tilde{\Sigma}_{pp}$ and -consequently-
$\tilde{\Sigma}_\Lambda$), blows up at low-frequencies. Hence, for
somewhat larger values of $U$ results similar to Figs.~\ref{fig:3.2}
are obtained. 
At the same time we find, consistent with
the findings of Ref.~[\onlinecite{divergence}], that {\sl no} vanishing
eigenvalue occurs in $\chi_{sp}(\nu,\nu', \omega=0)$ even at larger
$U$ and $\tilde{\Sigma}_{sp}$ remains well-behaved also at strong
coupling.

A more physical elaboration of the meaning of such a selective
appearance of singularities in the different channels will be given in
the last section of the paper.

Until now we have discussed the case $\omega=0$. For $\omega \ne 0$ there are 
negative diagonal matrix elements of $\chi_{ch}$ even for small values of $U$. 
For instance, already in a generalization of Eq.~(\ref{eq:5.1}) negative diagonal matrix elements can appear. These elements are 
particularly small for large $\nu$ and $\omega$. Hence, inverting such a matrix
gives large matrix elements for large $\nu$ and $\omega$, which are rather 
unimportant for the self-energy and, thus, not very interesting in the
light of the parquet decomposition.

\subsection{$N_c=4$ case}\label{sec:4.2}

\begin{table}[b!]
\caption{\label{table:5.1}Important matrix elements of 
$\chi_d({\bf K},\nu;{\bf K}',\nu';{\bf Q},\omega)$ for 
${\bf Q}=(\pi,\pi)$ and $\omega=0$. The parameters are 
$N_c=4$, $t=-0.25$ eV, $U=1.5$ eV and $\beta=10$ eV$^{-1}$.}
\begin{tabular}{rrrrrrrrrr}
\hline
\hline
$\nu$ & ${\bf K}$ & \multicolumn{4}{c}{$\nu'=-\pi/\beta$}& \multicolumn{4}{c}{$\nu'=\pi/\beta$} \\
\hline
 & & $(\pi,\pi)$ & $(\pi,0)$ & $(0,\pi)$ & $(0,0)$& $(\pi,\pi)$ & $(\pi,0)$ & $(0,\pi)$ & $(0,0)$ \\ 
\hline
$-\pi/\beta$ & $(\pi,\pi)$ & 7.4 & -1.7 & -1.7 &  -4.0 &   0.4 &   0.6 &  0.6 &  -0.0    \\ 
$-\pi/\beta$ & $(\pi,0)$ & -1.7  &   4.8     & -16.      &  -1.7     &   0.6     &   3.9     &   2.3     &   0.6      \\
$-\pi/\beta$    & $(0,\pi)$ &-1.7     & -16.      &   4.8     &  -1.7      &  0.6     &   2.3     &   3.9     &   0.6      \\ 
$-\pi/\beta$    & $(0,0)$ &-3.9     &  -1.7     &  -1.7      &  7.4     &   0.0     &   0.6     &   0.6     &   0.4     \\ 
$\pi/\beta$ & $(\pi,\pi)$ & 0.4     &   0.6     &   0.6     &   0.0     &   7.4     &  -1.7     &  -1.7     &  -4.0      \\
$\pi/\beta$ &  $(\pi,0)$ & 0.6     &   3.9     &   2.3     &   0.6     &  -1.7     &   4.8     & -16.      &  -1.7     \\
$\pi/\beta$& $(0,\pi)$ & 0.6     &   2.3     &   3.9     &   0.6     &  -1.7     & -16.       &  4.8     &  -1.7     \\
$\pi/\beta$& $(0,0)$ &  0.0     &   0.6     &   0.6     &   0.4      & -3.9     &  -1.7      & -1.7     &   7.4      \\
\hline
\hline
\end{tabular}
\end{table} 

We will now extend the previous DMFT analysis of the
singularities to the DCA calculations for $N_c=4$. 

In this case, $\chi_{ch}$ is also momentum dependent, and, in general,
a complex function.
However, at half-filling, for ${\bf Q}=(\pi,\pi)$ and $\omega=0$ it
remains purely real.\cite{noteotherQ} We therefore mostly focus on this case, which gives an important 
contribution to $\Sigma$. As in the previous section, we use the parameters $t=-0.25$ 
eV and $\beta=10$ eV$^{-1}$, and study the occurrence of vanishing
eigenvalues in $\chi_{ch}$.

Since, as discussed at the beginning of last section, we are not
interested in the high-frequency (perturbative) eigenvalues of
$\chi_{ch}$, we choose an interaction value, where the most important
fermion frequencies have already become the lowest ones: $\nu,\nu'=\pm \pi/\beta$.
In particular, Table~\ref{table:5.1} shows some of these matrix elements
for, e.g., $U=1.5$ eV. Here, one sees that the dominating off-diagonal 
matrix elements are obtained for $\nu=\nu'=\pm \pi/\beta$ and ${\bf K}\ne 
{\bf K}'$ taking values $(\pi,0)$ or $(0,\pi)$. Based on the size of
the different matrix elements 
in Table~\ref{table:5.1}, 
it is then natural to focus on the $4\times 4$ matrix containing the 
${\bf K}$-vectors $(\pi,0)$ and $(0,\pi)$ as well as the frequencies 
$\nu,\nu'=\pm \pi/\beta$ for ${\bf Q}=(\pi,\pi)$ and $\omega=0$. The 
lowest eigenvalue of this matrix is defined as $\varepsilon_{4 \times 4}$. 
We also calculate the lowest eigenvalue, $\varepsilon_{{\bf K} \times 
{\bf K}}$, corresponding to the the $2\times 2$ matrix containing the 
two ${\bf K}$-vectors at the Fermi-level, $(\pi,0)$ and $(0,\pi)$, for one frequency, 
i.e., $\nu= \pi/\beta$. Finally, we calculate the lowest eigenvalue, 
$\varepsilon_{\nu \times \nu}$ corresponding to the $2\times 2$ matrix 
containing two frequencies $\nu,\nu'=\pm \pi/\beta$ and one ${\bf
  K}=(\pi,0)$. 

\begin{figure*}
{\rotatebox{-90}{\resizebox{6.0cm}{!}{\includegraphics {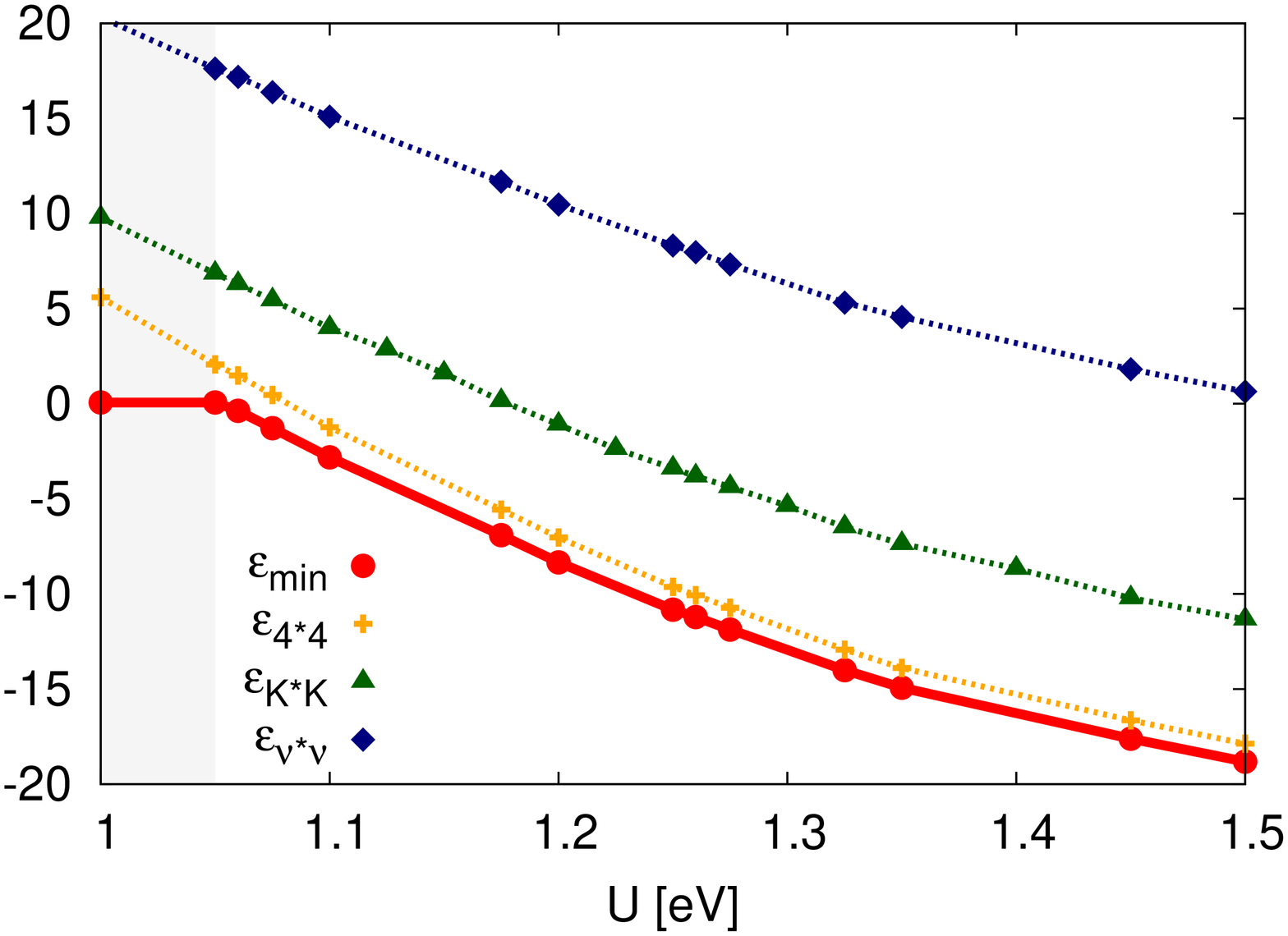}}}}
\hspace{1mm}
{\rotatebox{-90}{\resizebox{6.0cm}{!}{\includegraphics {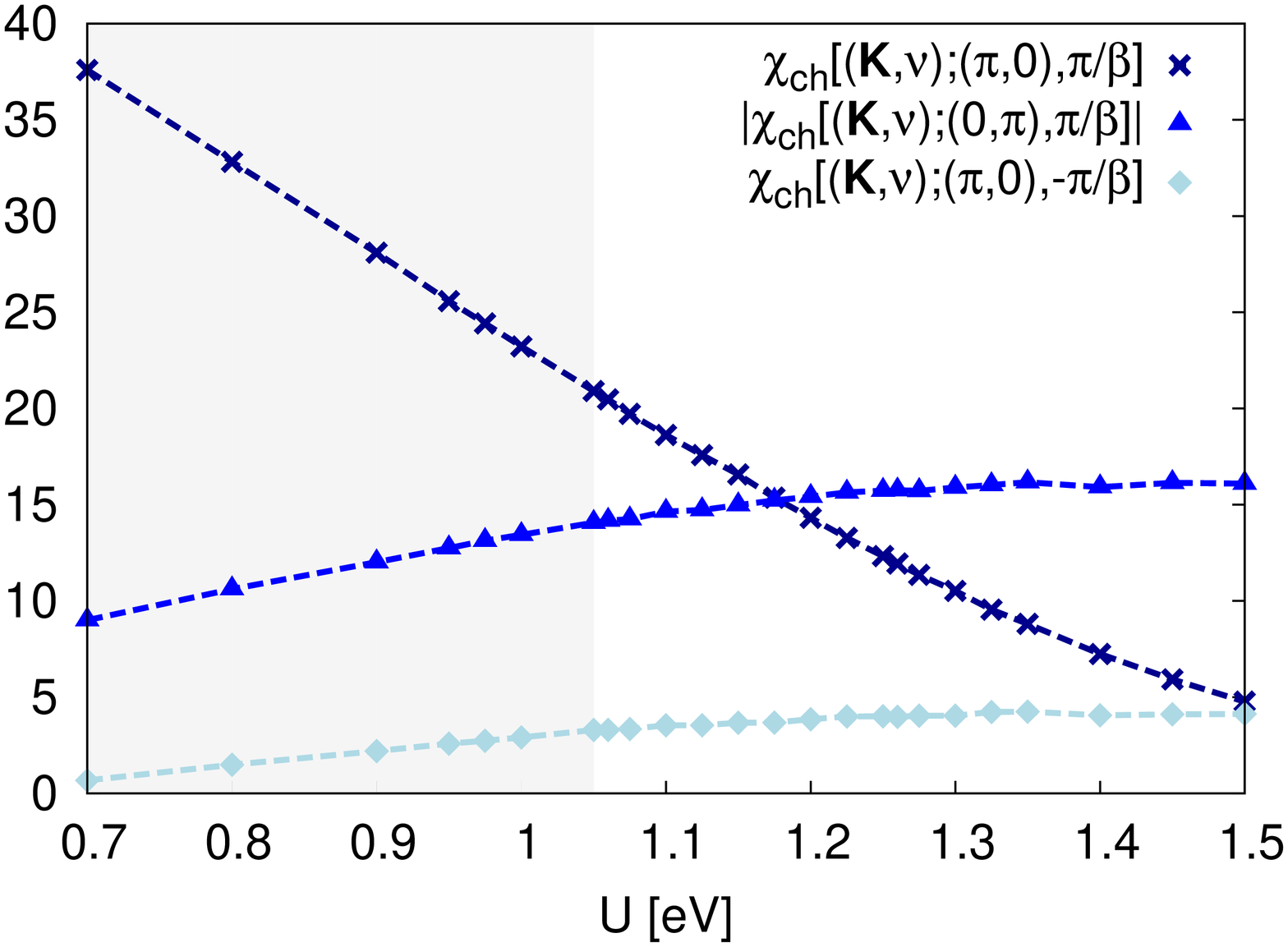}}}}
\caption{\label{fig:5.5}Lowest eigenvalue $\varepsilon_{\rm min}$  of $\chi_d$ 
compared with the approximations $\varepsilon_{4 \times 4}$, $\varepsilon
_{{\bf K}\times {\bf K}}$ and $\varepsilon_{\nu_n \times
  \nu_n}$. Matrix elements of $\chi_{ch}$ for ${\bf Q}=(\pi,\pi)$ and
$\omega=0$ as a function of U.The parameters 
are $N_c=4$, $t=-0.25$ eV eV and $\beta=10$ eV$^{-1}$.}
\end{figure*}

The results of our analysis, for different values of $U$, are shown in
Fig.~\ref{fig:5.5}. 
We see that the eigenvalue $\varepsilon_{4 \times 4}$
provides a quite accurate approximation to the exact minimal eigenvalue
$\varepsilon_{\rm min}$ of the full generalized charge susceptibility
for values of $U$ ($U > 1.05$) where $\varepsilon_{\rm min}<0$. This illustrates 
that the matrix elements discussed above are really the dominating
ones. Furthermore, we find that the ``Fermi-level''-momentum
approximation $\varepsilon_{{\bf K}\times {\bf K}}$ is also reasonably accurate, 
while the low-frequency $\varepsilon_{\nu \times \nu}$  is less accurate.

Fig.~\ref{fig:5.5} (right panel) shows the dependence on $U$ for some of these   
matrix elements. The diagonal element for ${\bf K}={\bf K}'=(\pi,0)$ 
and $\nu=\nu'= \pi/\beta$  rapidly decreases with $U$, while the 
absolute value of the off-diagonal element in ${\bf K}$ for ${\bf K}=(\pi,0)$, 
${\bf K}'=(0,\pi)$ and $\nu=\nu' =\pi/\beta$ is large and slowly 
increases with $U$. This matrix element is, in particular, due to 
the unequal spin contribution. In Sec.~\ref{sec:5} we show how  --
in the case of $N_c=4$ cluster -- this evolution  is  
linked to the progressive stabilization of a RVB-dominated ground-state. The off-diagonal element in 
frequency for $ {\bf K}={\bf K}'= (\pi,0)$, $\nu=\pi/\beta$ and 
$\nu'=-\pi/\beta$, instead, remains rather small.  

\begin{figure}
{\rotatebox{0}{\resizebox{6.0cm}{!}{\includegraphics {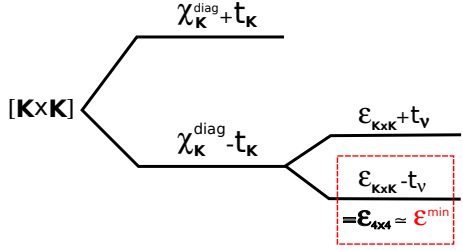}}}}
\caption{\label{fig:5.6} Schematic representation of the main
  diagonalization steps, determining the lowest ($1st$: dashed red box, and $2nd$)
  negative eigenvalues of $\chi_{ch}$, after the first/second level crossing has occurred.}
\end{figure}

The minimal eigenvalue $\varepsilon_{{\bf K} \times {\bf K}}$ of the $2\times 2$ matrix
in ${\bf K}$ is given by
\begin{eqnarray}\label{eq:5.14}
\varepsilon_{{\bf K} \times {\bf K}} &\!=\! & \chi_{ch}[(\pi\!,\!0),{\pi\over
  \beta};(\pi\!,\!0),{\pi \over \beta}] \!-\!
|\chi_{ch}[(\pi\!,\!0),{\pi\over\beta};(0\!,\!\pi),{\pi\over \beta}]|
\nonumber \\
  & \!=\! & \chi_{ch}^{diag} - t_{\bf{K}}
\end{eqnarray}
Evidently, when the magnitude of the off-diagonal element ($t_{\bf{K}}$) becomes equal to the diagonal
element ($\chi_{ch}^{diag}$), the lowest eigenvalue $\varepsilon_{{\bf  K} \times {\bf K}}$ goes negative (see
Fig.~\ref{fig:5.6}).  The (opposite) sign of the matrix elements in the $2
\times 2$ matrix are such that the two components of the corresponding eigenvector
have the {\sl same} sign. 

\begin{figure}
{\rotatebox{-90}{\resizebox{6.0cm}{!}{\includegraphics {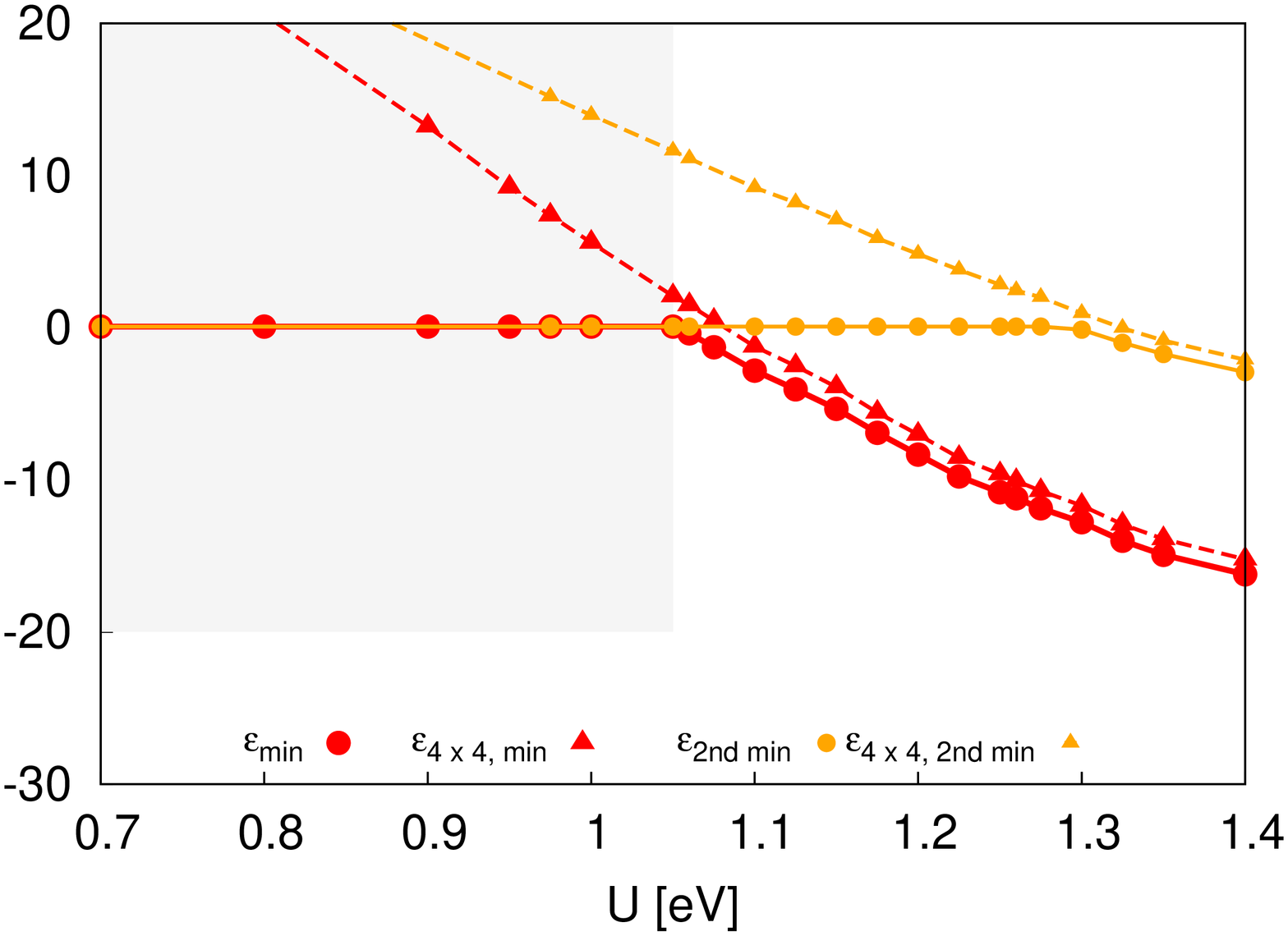}}}}
\caption{\label{fig:5.7} Calculated and approximate ($\varepsilon_{4 \times 4}$) 
lowest two eigenvalues. The parameters 
are $N_c=4$, $t=-0.25$ eV and $\beta=10$ eV$^{-1}$.}
\end{figure}

By extending our analysis, we will then the considered the  $4\times 4$
matrix. Its two lowest eigenvalues are shown in Fig.~\ref{fig:5.7}. The eigenvalue $\varepsilon_{{\bf K} 
\times {\bf K}}$ is further split into two by the small off-diagonal matrix elements for 
$\nu\ne \nu'$ ($t_{\nu}$ in Fig. ~\ref{fig:5.6}), in a bonding and
anti-bonding state. 
Similarly to the $N_c=1$ case the components of the eigenvector
corresponding to the lowest eigenvalue ($\varepsilon_{{\bf K} 
\times {\bf K}} - t_{\nu}$) have different signs for  $\nu=-\pi/\beta$
and $\pi/\beta$. 
Then, the eigenvector 
corresponding to the second lowest eigenvalue, which vanishes at a 
larger $U \sim 1.35$, will have the two component $\nu=\pm \pi/\beta$ 
with the same sign.

Eventually, combining all the eigenvector signs,  we obtain that the
lowest eigenvalue is associated to an eigenvector with opposite sign
components, while the second lowest is not. This evidently depends on
the specific signs in Table~\ref{table:5.1}.
 Hence, similar to the $N_c=1$ case,
also for $N_c=4$, the singularities occurring in $\chi_{ch}$ will
be actually responsible for the blowing up of the parquet decomposition (see Appendix \ref{appendix}), with
the significant exception of the first one
encountered from weak-coupling.

If one considered also the generalized susceptibility in
the particle-particle channel
$\chi_{pp,\uparrow\downarrow}$, we would find an analogous trend. For the case considered here where $\chi_{pp}$ is
real for ${\bf Q}=(0,0)$, it would show an eigenvalue going through
zero slightly below $U=1.3$ eV.  Similarly, the complex $\chi_{pp}$ for ${\bf Q}=(\pi,\pi)$ 
has a real eigenvalue going through zero slightly below $U=1.25$ eV. 
In such cases, the signs of the corresponding singular eigenvector
components do not compensate, which yield to the strong low-frequency
oscillations of the $\tilde{\Sigma}_{pp}$ data, presented in the previous section.
 Moreover, in the same parameter regime ($U \sim 1.3$), 
also the singularities of the lowest real eigenvalue of 
$\chi_{ch}$ for ${\bf Q}=(\pi,0)$ or for $(0,\pi)$ crossing zero do
not cancel, leaving the spin channel as the {\sl only} contribution of the
parquet decomposition unaffected by singularities.

In summary, we find that the singularities in the generalized
susceptibilities are actually reflected in a blowing up of the parquet
decomposition in the corresponding channel(s). Due to the possible
occurrence of compensating signs in the frequency components of the
singular eigenvector of $\chi$, however,
the correspondence is not complete. In fact, we find that the parquet decomposition remains well-behaved in
{\sl all} channels even beyond the value of $U$, where the first 
singularity appears in the Bethe-Salpeter equation for the charge
channel (i.e., at $U \sim 1.05$ for $\beta =10$ eV$^{-1}$ in DCA with $N_c=4$), because of the compensating
signs of the singular eigenvector.
However, this is no longer the case for larger values of $U$, where the singular parts 
of $\chi_{ch}^{-1}$ and/ or $\chi_{pp}^{-1}$ add up in the parquet decomposition of the 
self-energy, making a separate (parquet) treatment of the
corresponding scattering channels quite problematic. More specifically, in this regime, 
the absolute contribution from the totally irreducible diagrams to the self-energy 
at low-frequencies tends to be very large, and to a substantial extent, to be 
canceled by a very large particle-particle contribution. Beyond this compensations, 
it is also interesting to note that in Fig.~\ref{fig:3.3} a sign-crossing is 
observed between the anomalous low frequency contributions of the irreducible and 
the $pp$ channel to the self-energy and their more conventionally
behaved counter-parts at high-frequency.  Hence, since  the high-frequency behavior of the self-energy 
can be related to the lowest order perturbation theory, the sign crossings of the 
$pp$ and fully irreducible contribution  at intermediate frequency represent an 
evident manifestation of the break-down of the perturbative description. 

In order to go beyond this mostly formal interpretation of the
singularities in the generalized susceptibilities (and of their
effects on the parquet decomposition), in the next subsection we will
improve our understanding of the underlying physics by a comparison with
simplified model cases, where such singularities also appear.

\section{Physical interpretation of the singularities}\label{sec:5}

\subsection{Two level model}\label{sec:5.1}

To improve our physical insight on the occurrence of the
singularities, we start by considering one of the most basic case,
where they appear, i.e., a simple two-level (impurity) model:
This model has a Coulomb interaction $U$ on the ($N_c=1$) cluster site  
and no interaction on the bath site $b$ and an intersite hopping
$V$. Specifically,  we use $V=0.5$ eV, $\beta=5$ eV$^{-1}$ and we
consider the half-filled case. 

\begin{figure}
{\rotatebox{-90}{\resizebox{5.5cm}{!}{\includegraphics {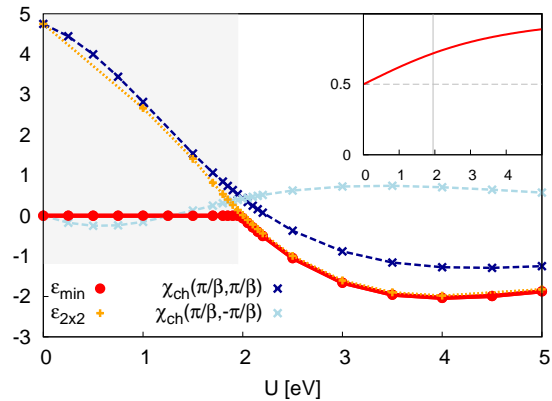}}}}
\caption{\label{fig:5.5a}
Diagonal and off-diagonal elements of $\chi_{ch}^{-1}$ for $\nu=\pm \pi/\beta$ 
of the two-level model as a function of $U$. The lowest eigenvalue is compared 
with the approximation in Eq.~(\ref{eq:5.2}) for the eigenvalue. The weight of 
the singlet component in Eq.~(\ref{eq:5.4}) is shown in the inset as a function 
of $U$. The parameters are $V=0.5$ eV and $\beta=5$ eV$^{-1}$.  
}
\end{figure}

Fig.~\ref{fig:5.5a}  shows the lowest eigenvalue $\varepsilon_{\rm min}$ of $\chi_{ch}$ for $\omega=0$ 
and the corresponding lowest eigenvalue $\varepsilon_{2 \times 2}$ in Eq.~(\ref{eq:5.2}) of 
the $2 \times 2$ matrix containing matrix elements for $\nu=\pm \pi/\beta$. 
More specifically we also note that increasing $U$ increases the off-diagonal matrix 
element $\chi_{ch}(\pi/\beta;-\pi/\beta)$. Similarly as for the DMFT
calculations of the previous section, when this element becomes equal 
to the diagonal element, $\varepsilon_{2 \times 2}$ goes negative [Eq.~(\ref{eq:5.2})]. 
At this point, $\varepsilon_{2 \times 2}$ becomes a rather good approximation 
to $\varepsilon_{\rm min}$, as it was also the case in Fig.~\ref{fig:5.2}. 
Hence, in this parameter range, we can limit our analysis to the lowest frequency sector ($\nu=\pm \pi/\beta$).

From the above discussion, we notice that the overall properties of the singularity of $\chi_{ch}$ in the two-level model
appear qualitatively similar to the one of the DMFT calculations of
the Hubbard model in Sec.~\ref{sec:4.1}.
Differently from the latter case, however, in the two-level model, we
have access to more intrinsic information, such as the exact ground state of
the systems. This allows for a deeper investigation of the physical evolution associated with the
singularities. In particular, we show how large the overlap of the ground state of the
system with the singlet state
\begin{equation}\label{eq:5.4}
{1\over \sqrt{2}}(|c\uparrow b \downarrow\rangle -|c \downarrow b\uparrow\rangle ),
\end{equation}
is, where two electrons, one on each site, form a valence bond: In the
inset of Fig.~\ref{fig:5.5a}, by increasing $U$, we clearly observe a
monotonously enhanced weight of the singlet state
of Eq.~\ref{eq:5.4} in the ground state of the system. 

In particular, the progressive change in the ground state is
responsible of the (increasing/decreasing) trends of the
(off-diagonal/diagonal) elements of $\chi_{ch}(\pm \pi/\beta; \pm
\pi/\beta)$, driving, eventually, the sign-change of 
$\varepsilon_{\rm min}$.

Below we will continue by  discussing the more significant $N_c=4$ case,
and show in more detail how the formation of a negative eigenvalue of $\chi_{ch}$
is, in that case, associated with the formation of a resonance valence bond (RVB).

\subsection{RVB state and pseudogap}\label{sec:5.2}

In this subsection, we will show how the analysis of properties of the
ground-state of the system can be extended to the case of
$N_c=4$. Here, instead of the two-level model, we will exploit  
a preceding study of the pseudogap in the Hubbard model using a
very different approach.\cite{pseudogap}  In fact, due to the relevance
for the cuprate physics, the general problem of the pseudogap formation in the
Hubbard model on a square lattice  has been intensively investigated for embedded
clusters, e.g., in DCA.\cite{otherpseudogapref,Leo}  Based on studies for $N_c=4$
and $8$, it has recently been argued that, for a sufficiently large $U$, a
{\sl localized} state $|{\rm \psi}_{\rm loc}\rangle $ is formed on the
cluster,\cite{pseudogap} leading to pseudogap features.  More specifically, by
comparing correlation functions of the DCA calculation and for $|{\rm \psi}_{\rm loc}\rangle$,
this state was identified\cite{pseudogap} with a singlet, which --for
$N_c=4$  we are considering here-- takes the approximate form
\begin{equation}\label{eq:3.1}
|{\rm \psi}_{\rm loc}\rangle \!= \!{1\over \sqrt{2}}(c^{\dagger}_{(\pi,0)\uparrow}c^{\dagger}_{(\pi,0)\downarrow}-
c^{\dagger}_{(0,\pi)\uparrow}c^{\dagger}_{(0,\pi)\downarrow})c^{\dagger}_{(0,0)\uparrow}
c^{\dagger}_{(0,0)\downarrow}|{\rm vac}\rangle
\end{equation}
Here, the ${\bf K}=(0,0)$ level is doubly occupied, while the levels ${\bf K}=(\pi,0)$
and $(0,\pi)$ are each doubly occupied with a probability of $\frac{1}{2}$.
We now want to show that this state is closely related to the resonance valence bond
(RVB) state.\cite{Liang88} Since the RVB
state has no double-occupancy ($U \rightarrow \infty$), we can make this connection
explicit in two steps. First we compare with a calculation for an isolated cluster
with $t=-0.25$eV and a finite, intermediate value of $U=1.25$ eV, relevant for the discussion here.
Afterwards, we compare these calculations for the isolated cluster with $U=1.25$ eV and $U=\infty$.
We find a very large overlap ($\sim 0.92$) between $|{\rm \psi}_{\rm loc}\rangle$
of Eq.~(\ref{eq:3.1}) and the ground-state of the isolated $U=1.25$ eV
cluster. Secondly, we find that the overlap of the ground-state for the isolated
cluster with $U=1.25$ eV to the $U=\infty$ RVB state is also large ($\sim 0.85$),
the difference arising mainly from the double-occupancies. In fact, all configurations
in real space with nonzero weight for the RVB state have similar weights
also in the calculation for $U=1.25$ eV. In summary, $|{\rm \psi}_{\rm loc}\rangle $
in Eq.~(\ref{eq:3.1}) is closely related to the ground-state of the isolated cluster
at finite $U$, and, hence, apart from some residual double occupancy, to the RVB state.

\begin{figure}
{\rotatebox{-90}{\resizebox{6.0cm}{!}{\includegraphics {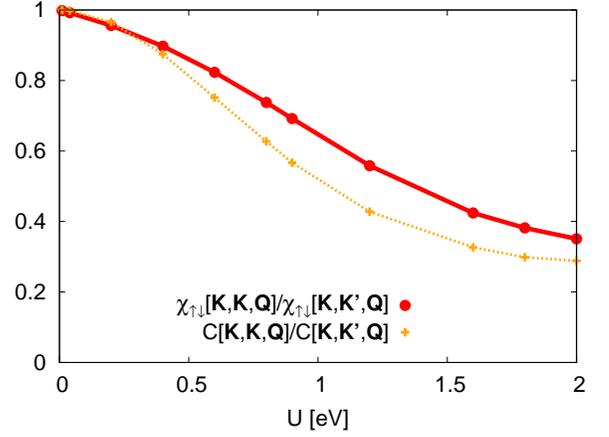}}}}
\caption{Ratio of $\chi_{\uparrow \downarrow}^{\bf K, K, Q}/\chi_{\uparrow \downarrow}^{{\bf K},{\bf K^\prime},{\bf Q}}$ for
${\bf K}=(0,\pi)$ and ${\bf K}'=(\pi,0)$ and ${\bf Q}=
(\pi,\pi)$. The figure also shows the ratio $C[{\bf K}, {\bf K}, {\bf Q}]/C[{\bf K}, {\bf K}', {\bf Q}]$ (Eq.~\ref{eq:3.0a}).  Matsubara labels have been suppressed; all Matsubara frequencies have their $n=0$ values. The parameters are 
$t=-0.25$ eV, $\beta=10$ eV$^{-1}$ and $N_c=4$. 
\label{fig:5.2}
}
\end{figure}

We now want to show that the state in Eq.~(\ref{eq:3.1}) is indeed formed and to 
relate this to the divergence of $\chi_{ch}$. We focus on the case $N_c=4$. 
As discussed in the context of Fig.~\ref{fig:5.7}, an important reason for the
divergence is the behavior of $\chi_{ch}$ and in particular of $\chi_{\uparrow \downarrow}(k,k',q)$ for ${\bf Q}
=(\pi,\pi)$ and ${\bf K}$ and ${\bf K}'$ equal to $(\pi,0)$ or $(0,\pi)$ at the 
lowest Matsubara frequencies. As $U$ is increased the element for ${\bf K}={\bf K}'$ 
is reduced while the element for ${\bf K}\ne{\bf K}'$ becomes large and negative.
To make the connection between the formation of an RVB state and the divergence,
we introduce
\begin{equation}\label{eq:3.0a}  
C({\bf K},{\bf K}',{\bf Q})=\sum_{\nu \nu'\omega}\chi_{\uparrow\downarrow}
({\bf K},\nu;{\bf K}',\nu';{\bf Q},\omega).
\end{equation}
Fig.~\ref{fig:5.2} shows that the ratio between $C$ for ${\bf K}'={\bf K}$ and
${\bf K'}\ne {\bf K}$ behaves in a very similar way as the corresponding ratio
for $\chi_{\uparrow\downarrow}({\bf K}, \pi/\beta;{\bf K}', \pi/\beta;{\bf Q},0)$
at the lowest Matsubara frequencies. The difference between the two curves is 
that $C$ contains a sum over all Matsubara frequencies. It is then not surprising 
that the two curves are similar. The quantity in Eq.~(\ref{eq:3.0a}) is 
easier to analyze. We use that $(1/\beta)\sum_{\nu}e^{i\tau \nu}=\delta(\tau)$,
where the summation is over fermion or boson frequencies. Then
\begin{equation}\label{eq:3.3}
{1\over \beta^3}C({\bf K},{\bf K}',{\bf Q})=\langle c^{\dagger}_{{\bf K}\uparrow}
c^{\phantom \dagger}_{{\bf K}+{\bf Q}\uparrow} c^{\dagger}_{{\bf K}'+{\bf Q}\downarrow} 
c^{\phantom \dagger}_{{\bf K}'\downarrow}\rangle
\end{equation}
It is then easy to check that for the ground-state (\ref{eq:3.1}) the matrix element
for ${\bf Q}=(\pi,\pi)$, ${\bf K}={\bf K}'=(\pi,0)$ is zero, while it is $-\frac{1}{2}$ for ${\bf K}'=(0,\pi)$.
This would lead to a vanishing ratio in Fig.~\ref{fig:5.2}, in qualitative agreement with the
actual calculation. 

The second lowest state on the cluster is a triplet of the form
\begin{equation}\label{eq:3.4}
{1\over
  \sqrt{2}}(c^{\dagger}_{(\pi,0)\uparrow}c^{\dagger}_{(0,\pi)\downarrow} +
c^{\dagger}_{(0,\pi)\uparrow}c^{\dagger}_{(\pi,0)\downarrow})c^{\dagger}_{(0,0)\uparrow}c^{\dagger}_{(0,0)\downarrow}|{\rm vac}\rangle.
\end{equation}
It should be emphasized here, that if this had been the lowest state,
we would have got exactly the {\sl opposite} result to above, i.e.,
a large matrix element for ${\bf K}={\bf K}'$ and a small matrix
element for ${\bf K}'=(0,\pi)$.

Our analysis of the $N_c=4$ DCA results demonstrate thus that in the regime, where
a pseudogap is observed\cite{pseudogap} for sufficiently large $U$,
i) the essential physics can be traced back to a state of RVB character,
and  ii) that the hallmark of such RVB character is directly reflected
in large off-diagonal elements of $\chi$ in the Fermi-momentum
subspace [$(0,\pi)$, $(\pi,0)$] . The latter result is quite
important for discussing the interpretation of the observed singularities in the
parquet decomposition of the DCA results. At large enough $U$, in
fact, an underlying 
RVB state has also been related to the formation of a pseudogap.\cite{pseudogap} 
Thus, in this regime, the trends towards a RVB ground-state would be the common
underlying reason behind 
onset a pseudogap and the formation of negative eigenvalues
of $\chi_{ch}$ and the associated strong frequency oscillations of the
parquet decomposition.

We note, finally, that the considerations discussed here are
rigorously valid for the parquet singularities of the $N_c=4$ data. They will
remain largely applicable to the cases of small DCA clusters discussed
in the present work. Modifications might be possible, instead, in the cases of extended
clusters, where a pseudogap spectral weight suppression can be
induced also at much weaker-coupling by long-ranged (spin)
correlations\cite{Schaefer2015A}. For such larger DCA clusters, the parquet
decompositions is still numerically challenging.

\subsection{Charge susceptibility and closeness to Mott transition}\label{sec:5.3}

Some further physical insight into this problem can be gained starting from the 
general observation that, when $U$ is increased, the charge susceptibility
 is suppressed, while the spin susceptibility becomes large. It is then not 
surprising that we find rather different behavior of $\chi_{ch}$ and $\chi_{sp}$. 
The charge susceptibility can be expressed in terms of the generalized charge 
susceptibility
\begin{equation}\label{eq:6.1}
\chi_{ch}(q)={1\over N_c\beta^2}\sum_{kk'}\chi_{ch}(k;k';q).
\end{equation}
We now use Eq.~(\ref{eq:5.3}) to rewrite the susceptibility as
\begin{equation}\label{eq:6.2}
\chi_{ch}=\sum_i \sum_{kk'} \langle k|i\rangle \varepsilon_i \langle i|k'\rangle
=\sum_i \varepsilon_i |\sum_k \langle k|i\rangle|^2,
\end{equation}
where $\varepsilon_i$ and $|i\rangle$ are the eigenvalues and eigenvectors, respectively,
of $\chi_{ch}$. The $q$ dependence is not shown explicitly. We find
\begin{equation}\label{eq:6.3}
\sum_i |\sum_k \langle k|i\rangle|^2=\sum_{kk'}\sum_i \langle k|i\rangle \langle i|k'\rangle=N_k,
\end{equation}
where $N_k$ is the number of $k$ values and thereby the number of
eigenvalues.  Thus, except ``pathological'' cases of strongly varying
overlaps $\langle k|i\rangle$ occur\cite{note_evsdem}, $|\sum_k \langle k|i\rangle|^2$  will be in
general {\sl not} small. This, together with $\chi_d$ being 
small, puts then  constraints on the eigenvalues.

For $N_c=1$ it has been shown that all eigenvalues of $\chi_{ch}(k,k',q=0)$ are positive 
for small $U$.\cite{divergence}  For large $U$, a small $\chi_{ch}(q=0)$ can be obtained 
if all eigenvalues are small (and possibly all positive) or if some eigenvalues are 
negative. Since individual matrix elements are large, the former could not be the case. 
Then the strong suppression of $\chi_{ch}(q=0)$ for large $U$ is expected to require that 
some eigenvalues are negative, although pathological cases may be found where this is 
not the case.  Similar arguments apply for larger clusters for values of $q$ where 
the eigenvalues are real and positive for small $U$. The appearance of negative 
eigenvalues as $U$ is increased and, hence, of the huge low-frequency oscillations in the the parquet decomposition,
should be a consequence of a gradual suppression of charge fluctuation as the system
approaches a Mott transition.  This supports an earlier preliminary interpretation
(within DMFT) of a negative eigenvalue as a precursor effect of the Mott 
transition.\cite{divergence} The DCA results, suggesting an intrinsic connection with
the RVB physics and the pseudogap formation, implies a more profound, and highly
non-perturbative, picture of the electronic correlations in
two-dimensional lattice systems.

\section{Conclusions}\label{sec:6}

We have calculated the two-particle vertex function in DMFT and  DCA for the
Hubbard model. The vertex function was then exploited to
perform a parquet decomposition of  the DCA self-energy. 
The purpose of such decomposition was similar as for the recently
introduced fluctuation diagnostic approach,\cite{FluctDiag} i.e., to improve our understanding of the 
physical origin of the numerical results for the self-energy. In comparison to
the latter approach, the parquet decomposition allows -in principle- for a more direct
formulation, which does not require any representation change in the
equation of motion for the self-energy. However, as we discussed in
this work, as opposed to the fluctuation diagnostics procedure, its usage poses also
important new challenges.

While the parquet decomposition works relatively smoothly in the
perturbative regime and allows one to evaluate quantitatively the role played by the
different channels, for larger values of 
$U$ and moderate doping, some of its terms start to display very
large oscillations at small frequencies. This renders it impossible to
disentangle the role of the channels affected by such oscillations. 
We should note, however, that in all cases considered we could
always find, even at strong coupling, at least {\sl one} well-behaved term in the parquet
decomposition (which was the spin contribution, $\tilde{\Sigma}_{sp}$, for the $2d$
and $3d$ Hubbard model close to/at half-filling). This has been
interpreted as a specific indication emerging from the parquet
decomposition of a predominance of that well-behave channel.  In this way the
predictions of the parquet decomposition of $\Sigma$ provide a qualitatively similar
outcome\cite{FluctDiag} to those of the fluctuation diagnostics.
 Unlike the former, the latter approach, appears not to be
 affected at all by entering in non-perturbative regime.

Beyond the physical insight in the self-energy, our results are
also relevant for the future developments of forefront methods in
quantum many body physics. In fact, several recently proposed 
computational schemes have been based on the parquet decomposition,
introducing approximations for the totally irreducible diagrams,
and then calculating the reducible diagrams via the parquet
equations.\cite{PA_Jarrell,DGA,Multiscale,DFparquet} The results above,
however, show that the contribution from the irreducible diagrams
becomes highly complicated for strongly correlated systems, even diverging for
certain values of $U$. This suggests that all schemes based on the
parquet decomposition above might encounter unforeseen problems
in the intermediate-to-strong correlated regime. However, we should recall that the generalized
susceptibilities in Matsubara space are not directly measurable
quantities. Hence, one may wonder, whether 
alternatives to the conventional parquet decomposition for classifying
the Feynman diagrams could be found, in order to improve the description of electronic correlations
in the intermediate coupling regimes and avoid the singularities.

In the specific context of our DMFT and DCA analysis, we have demonstrated  
that the singularities of some terms of the parquet decomposition of the self-energy is 
directly related to the divergencies of $\chi_{ch}^{-1}$ and $\chi_{pp}^{-1}$ at 
intermediate $U$ values.  In particular, we showed that the divergence of $\chi_{ch}^{-1}$ is 
related to the suppression of charge fluctuations. This represents an early, 
non-perturbative, manifestation of the Mott-Hubbard physics. The relation of 
such singularities to a RVB state and to the formation of a pseudogap has also
been investigated for the case of the $N_c=4$ DCA clusters, making
progress towards a theoretical understanding 
of the highly non-trivial physics of strong electron correlations in 
two-dimensions.

\section{ACKNOWLEDGMENTS}\label{sec:7}
We thank E. Gull and P. Thunstr{\"o}m for insightful discussions. We acknowledge financial support
from the Austrian Science Fund (FWF) through the Doctoral School Solid
for Fun (W1243, T.S.) the project I610-N16 (T.S,G.R.), and the
SFB-ViCoM F41 (A.T.). J.M. acknowledges financial support from MINECO (MAT2012-37263-C02-01).~G.S. acknowledges financial support 
from research  unit  FOR  1346  of  the  Deutsche Forschungsgemeinschaft. 
We also thank K. K\"olbl for graphical advices.

\appendix

\begin{figure*}[ht!]
{\rotatebox{-90}{\resizebox{5.7cm}{!}{\includegraphics {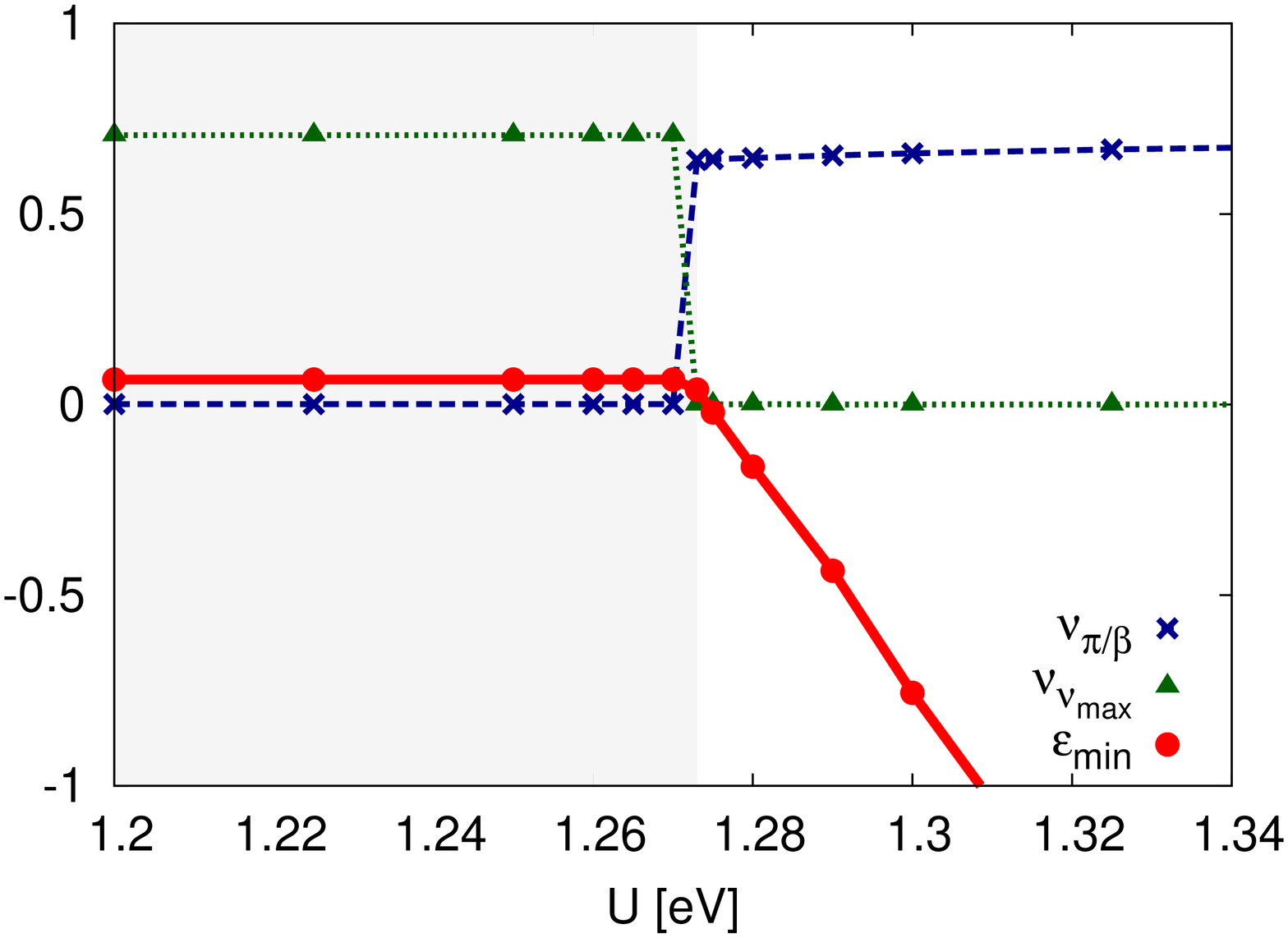}}}}
\hspace{5mm} 
{\rotatebox{-90}{\resizebox{5.7cm}{!}{\includegraphics {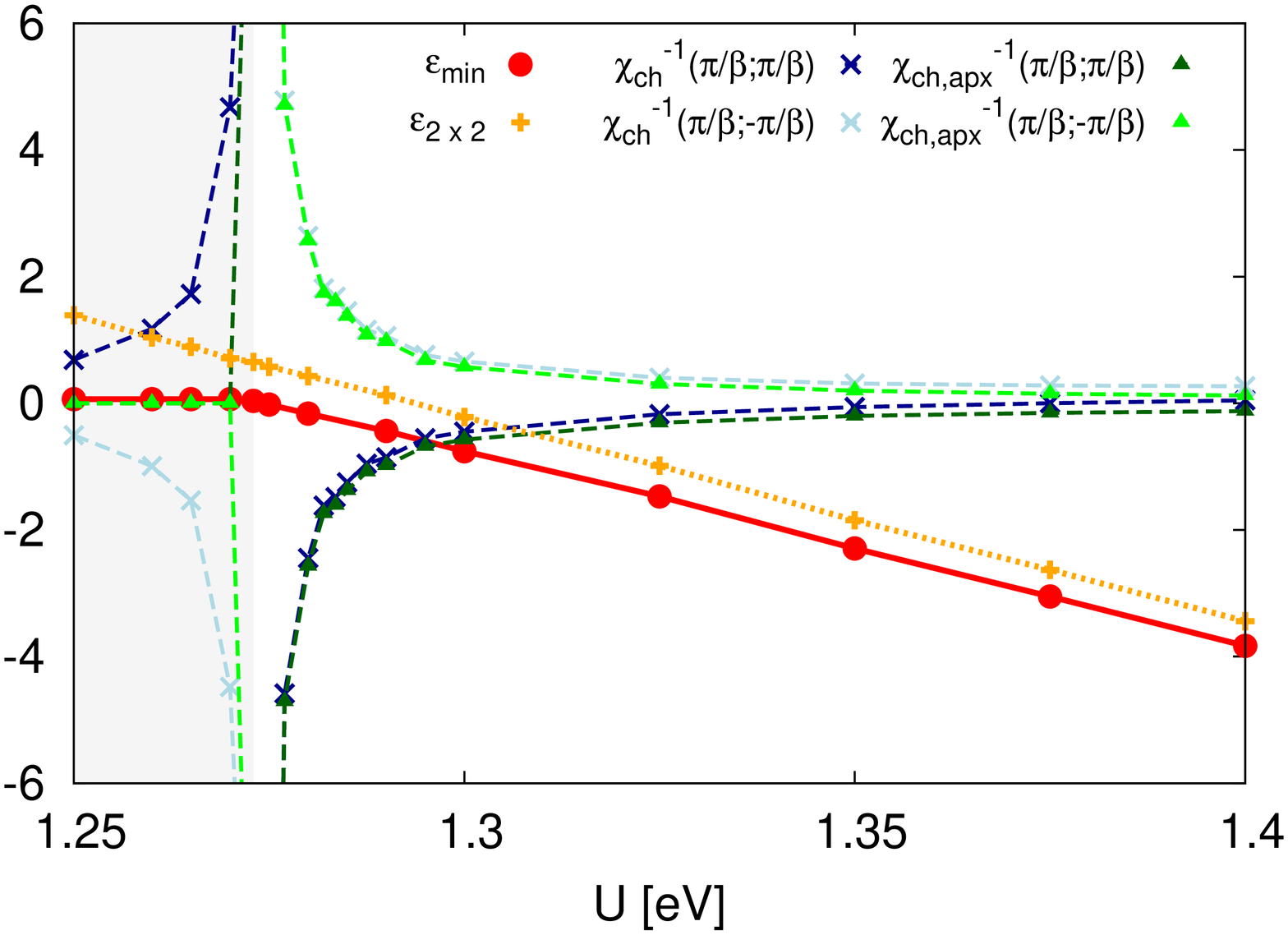}}}}
\caption{\label{fig:5.3} DMFT results illustrating the change of sign of the eigenvalues of $\chi_{\text{ch}}$ for the
parameters $t=-0.25$ eV, $\beta=10$ eV$^{-1}$, $n=1$ and $N_c=1$. Left panel: The lowest eigenvalue of $\chi_{\text{ch}}(\nu,\nu'\omega=0)$ and
the elements of the corresponding eigenvector for $\nu = \pm \pi/\beta$ and $\nu = \pm \nu_{\text{max}}$ are compared (see also text below Eq. (\ref{eq:5.1})). Right panel: Diagonal and off-diagonal elements of $\chi^{-1}_{\text{ch}}$ for $\nu=\pm \pi/\beta$. $\chi_{\text{ch,apx}}^{-1}$ 
is an approximation to $\chi_{\text{ch}}^{-1}$, using only the lowest eigenvalue in Eq.~(\ref{eq:5.3}).}
\end{figure*}

\section{Formation of negative eigenstate at $\omega=0$}\label{appendix}

\begin{figure}
{\rotatebox{-90}{\resizebox{6.0cm}{!}{\includegraphics {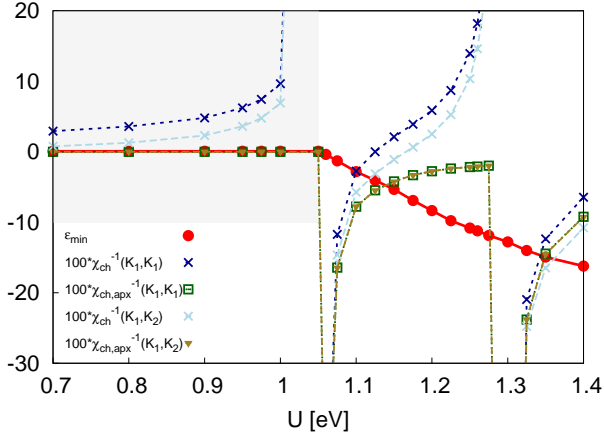}}}}
\caption{\label{fig:5.9} Calculated ($\chi_{\text{ch}}^{-1}$) and approximated ($\chi_{\text{ch,apx}}^{-1}$) 
matrix elements in DCA for the parameters $N_c=4$, $t=-0.25$ eV, $n=1$ and $\beta=10$ eV$^{-1}$. Shown are the two momentum points $K_1=(0,\pi)$ and $K_2=(\pi,0)$.}
\end{figure}

In this Appendix we further elaborate on the divergence of $\chi_{\text{ch}}^{-1}$ at $\omega=0$ in Sec. \ref{sec:4} and the corresponding evolution of the 
singular eigenvalue of $\chi_{\text{ch}}$. Specifically, in order to analyze the role played by the lowest eigenvalue of the generalized susceptibility, we express the inverse of $\chi_{\text{ch}}$ in the basis of the eigenvalues ($\varepsilon_i$) and the eigenvectors
($|i\rangle$) of $\chi_{\text{ch}}$:
\begin{equation}\label{eq:5.3}
\chi_{\text{ch}}^{-1}=\sum_i |i\rangle \varepsilon_i^{-1} \langle i|.
\end{equation}
An approximate expression, $\chi_{\text{ch,apx}}^{-1}$, can then be 
obtained by restricting the sum to the lowest eigenvalue of $\chi_{\text{ch}}$. 

We now illustrate the usefulness of the representation Eq. \ref{eq:5.3} by applying it first to the case of DMFT ($N_c=1$). In 
Fig.~\ref{fig:5.3}, the evolution of the exact and approximate eigenvalues with interaction strength $U$ is shown for $t=-0.25$ eV, $n=1$ and $\beta=10$ eV$^{-1}$.
 For $U<1.275$, where the
 lowest eigenvalue is positive, the contribution for $\nu=\pm \pi/\beta$ 
is approximately  zero, because the corresponding (weak-coupling) eigenvector 
has almost no weight for these frequencies. $\chi_{\text{ch}}^{-1}$ 
becomes large already for $U$ slightly smaller than $1.275$, where the 
approximate eigenvalue is small but positive. Here, a low-lying ``resonance'' gives a large contribution. When the resonance 
goes through zero and becomes a ``bound state'' (negative eigenvalue)
for the matrix of the generalized susceptibility, the sign of $\chi_{\text{ch}}^{-1}$ 
(and hence also that of $\Gamma_{\text{ch}}$) changes. For $U\ge 1.275$, $\chi_{\text{ch,apx}}^{-1}$ 
provides a quite good approximation of $\chi_{\text{ch}}^{-1}$, showing that the large 
values of $\chi_{\text{ch}}^{-1}$ are mainly due to this ``bound-state''.  As $U$ is increased further, 
the lowest eigenvalue gets more negative, and the matrix elements of 
$\chi_{\text{ch}}^{-1}$ are reduced. The basic character of $\chi_{\text{ch}}$, however, 
remains qualitatively different compared with smaller values of $U$.  

This analysis can also be extended to the case of DCA. 
Fig.~\ref{fig:5.9} shows matrix elements of the DCA $\chi_{\text{ch}}^{-1}$ (at half-filling and $\beta=10.0$) compared
 with  the approximation $\chi_{\text{ch,apx}}^{-1}$ where only {\sl negative}
 eigenvalues are  considered in the inversion in
 Eq.~(\ref{eq:5.3}). For $U<1.05$ all eigenvalues are are positive and $\chi_{\text{ch,apx}}^{-1}$ is zero. However, there is  a
 resonance for $U$ close to 1.05, as is also indicated by the small
 value  of $\varepsilon_{4 \times 4}$ (see Sec. \ref{sec:4} for the corresponding definitions). This leads to a large
 contribution to   $\chi_{\text{ch}}^{-1}$ for $U$ close to 1.05. As the lowest
 eigenvalue goes negative at $U=1.05$, the signs of some large matrix
 elements of $\chi_{\text{ch}}^{-1}$ change [see Eq.~(\ref{eq:5.3})]. At the
 same time $\chi_{\text{ch,apx}}^{-1}$ becomes a rather  good approximation to
 $\chi_{\text{ch}}^{-1}$. Increasing $U$ further, a second resonance  forms, as
 is also seen by the small value of the second lowest eigenvalue in
 the $ 4\times 4$ space. This leads to very large values of $\chi_{\text{ch}}^{-1}$ for 
 $U>1.2$, which are missed by $\chi_{\text{ch,apx}}^{-1}$.  For $U>1.35$ this resonance 
 is converted to a negative eigenvalue, signs of matrix elements of $\chi_{\text{ch}}^{-1}$ 
change, and $\chi_{\text{ch,apx}}^{-1}$ again becomes a good approximation of $\chi_{\text{ch}}^{-1}$.
  
\vspace{5mm}

\section{General structure of the $2\times 2$ singular matrix}  
\label{app:2by2}

The following generic matrix is related to the discussion in the main
text:

\begin{equation}
 \label{equ:2tim2}
 M=\begin{pmatrix}
	 a & b \\
	 b & a
 \end{pmatrix},
\end{equation}
where $a,b\in\mathds{R}$ and $b>0$. 

The eigenvalues and eigenvectors are given by $\lambda_{\mp}=a\mp b$
and $\mathbf{v}_{\mp}=(\mp 1,1)/\sqrt{2}$. 

Hence, the spectral representation of the inverse of $M$ reads:
\begin{equation}
\label{equ:minv}
M^{-1}=\frac{1}{2(a-b)}\begin{pmatrix}
	1 & -1 \\ -1 & 1
\end{pmatrix}+\frac{1}{2(a+b)}\begin{pmatrix}
	1 & 1 \\ 1 & 1
\end{pmatrix}
\end{equation}

When the first eigenvalue vanishes ($a=b$) the first term of the
matrix $M^{-1}$ diverges, while the sum over all its matrix
elements stays finite, because the sum over the matrix elements in the
first term exactly vanishes due to the antisymmetry of the
corresponding eigenvector. Hence, the sum over all matrix elements,
originating from the second term in Eq. (\ref{equ:minv}), yields the
finite result $1/a$. For $a=-b$, however, one encounters the divergence
of the second eigenvalue. In this case also the sum over all matrix
elements diverges, as the (equal) signs of the corresponding eigenvector no
longer cancel it.


\begin{thebibliography}{30}
\bibitem{notelowT} We note that, by reducing $T$ the weight of this
  eigenvector will become distributed over several low-lying
  frequencies, but the general picture, as well as the qualitative
  difference to   the weak-coupling regime,  
remains similar in the whole temperature interval we considered.
\bibitem{note_evsdem} It is worth stressing, that -- in the light of
  the recent results obtained via the fluctuation diagnostics\cite{FluctDiag}- the
  occurrence of strongly varying matrix elements looks unlikely to
  be realized in secondary scattering channels, such as the
  charge channel in our case. Rather, the secondary nature of the (here: charge) channel 
would be reflected in a relatively {\sl featureless} and smooth matrix
element behavior of the (corresponding) generalized susceptibility.
\end{thebibliography}
\end{document}